\documentclass[lettersize,journal]{IEEEtran}
\usepackage{cite}
\usepackage{bookmark}
\usepackage{subfig}
\usepackage[pdftex]{graphicx}
\usepackage[justification=centering]{caption}

\usepackage{amsmath}
\usepackage{amssymb}
\usepackage{bbm}
\usepackage{textcomp}
\usepackage{mathtools}
\usepackage{multirow}
\usepackage{stfloats}
\DeclareCaptionLabelFormat{simple}{(#2)}
\usepackage{cuted}
\usepackage{bm}
\usepackage{booktabs}

\usepackage{algorithm}
\usepackage{algorithmic}
\usepackage{chngcntr}

\usepackage{flushend}

\usepackage{comment}

\usepackage{tikz}
\usetikzlibrary{decorations.pathreplacing,arrows.meta,positioning}

\usepackage{pgfplots}
\usepackage{pgfplotstable}
\usepgfplotslibrary{groupplots}
\pgfplotsset{compat=1.18}

\pgfplotsset{
  discard if not/.style 2 args={
    x filter/.code={
      \edef\rowval{\thisrow{#1}}%
      \edef\target{#2}%
      \ifx\rowval\target\else
        \def\pgfmathresult{}%
      \fi
    }
  }
}

\newcommand{\rom}[1]{%
  \textup{\lowercase\expandafter{\romannumeral#1}}%
}

\DeclarePairedDelimiter\norm{\lVert}{\rVert}

\floatname{algorithm}{Algorithm}

\begin{document}

\title{Neural CSI Compression Fine-Tuning: Taming the Communication Cost of Model Updates}

\author{Mehdi~Sattari,~\IEEEmembership{Graduate~Student~Member,~IEEE,}
        Deniz~Gündüz,~\IEEEmembership{Fellow,~IEEE,}
        and
        Tommy~Svensson,~\IEEEmembership{Senior~Member,~IEEE}
        ~
\thanks{M.~Sattari, and T.~Svensson are with the Department
of Electrical Engineering, Chalmers University of Technology, Gothenburg,
Sweden. E-mail: \{mehdi.sattari, tommy.svensson\}@chalmers.se. 
D.~Gündüz is with the Department of Electrical and Electronic Engineering, Imperial College London, London,
UK. E-mail: d.gunduz@imperial.ac.uk.}
\thanks{This work has been supported in part by the project SEMANTIC, funded by the EU’s Horizon 2020 research and innovation programme under the Marie Skłodowska-Curie grant agreement No 861165, and in part by the Hexa-X-II project which has received funding from the Smart Networks and Services Joint Undertaking (SNS JU) under the European Union’s Horizon Europe research and innovation programme under Grant Agreement No 101095759. D. Gündüz received funding from the UKRI through ERC consolidator project AI-R under grant EP/X030806/1, and from the Horizon Europe SNS project ‘6G-GOALS’ under grant 101139232.
The computations were enabled by resources provided by the National Academic Infrastructure for Supercomputing in Sweden (NAISS), partially funded by the Swedish Research Council through grant agreement no. 2022-06725.}
\thanks{Parts of this work were presented at the IEEE International Workshop on Signal Processing and Artificial Intelligence for Wireless Communications (SPAWC), 2025 \cite{11143260}.}
}

\maketitle

\begin{abstract}
Efficient channel state information (CSI) compression is essential in frequency division duplexing (FDD) massive multiple-input multiple-output (MIMO) systems due to the substantial feedback overhead. Recently, deep learning-based compression techniques have demonstrated superior performance for CSI feedback. However, their performance often degrades under distribution shifts across wireless environments, largely due to limited generalization capability.
To address this challenge, we consider a full-model fine-tuning scheme, in which both the encoder and decoder are jointly updated using a small number of recent CSI samples from the target environment. A key challenge in this setting is the transmission of updated decoder parameters to the receiver, which introduces additional communication overhead.
To mitigate this bottleneck, we explicitly incorporate the bit rate of model updates into the fine-tuning objective and entropy-code the model updates jointly with the compressed CSI. Furthermore, we employ a structured prior that promotes sparse and selective parameter updates, thereby significantly reducing the model-update communication cost.
Simulation results across multiple CSI datasets demonstrate that full-model fine-tuning substantially improves the rate–distortion performance of neural CSI compression, despite the additional cost of model updates. We further analyze the impact of the evaluation horizon, the quantization resolution of model updates, and the size of the target-domain dataset on the overall feedback efficiency.
\end{abstract}


%
\IEEEpeerreviewmaketitle

\section{Introduction}
\IEEEPARstart{M}{assive} multiple-input multiple-output (MIMO) is one of the key enabling technologies for 5G and beyond. By equipping the base station (BS) with massive antenna arrays, a large number of users can be served simultaneously through high-resolution beamforming techniques, resulting in remarkable spectral efficiency. To enable massive MIMO technology, channel state information (CSI) must be available at both the BS and user sides. Various channel estimation techniques can be employed to estimate CSI by observing pilot signals. In time-division duplexing (TDD), only downlink/uplink channels need to be estimated, as the other channel can be derived thanks to reciprocity. However, in frequency-division duplexing (FDD), reciprocity does not hold, so the channels in both directions need to be estimated and fed back\cite{5595728}, \cite{6736761}.

To enable FDD transmission for massive MIMO systems, CSI compression techniques are critical to circumvent excessive feedback overhead. Various CSI compression techniques have been studied in the literature by employing compressed sensing \cite{6816089}, \cite{7442899}, vector quantization \cite{10208156}, \cite{6497019}, and more recently, deep learning tools \cite{8322184, 8972904, 9149229, 9797871, 9705497, 10130108, wu2024mimochannelneuralfunction, 9296555, chen2023csipppnetonesidedoneforalldeep}. Due to strong spatial correlation, the CSI matrix can exhibit sparsity in certain domains, and compressed sensing methodologies can be applied to efficiently compress the large CSI matrix \cite{6816089}, \cite{7442899}. However, these algorithms often struggle to find the best basis to project the CSI matrix to lower dimensions. Furthermore, compressed sensing-based algorithms are iterative and time-consuming, making them infeasible for deployment in massive MIMO channels. Similarly, in vector quantization for CSI compression \cite{10208156}, \cite{6497019}, the overhead scales linearly with the channel dimension, rendering it impractical for massive MIMO systems.


Recent advances in data-driven compression approaches, driven by the progress in neural network (NN) architectures have gained substantial interest in all data modalities \cite{10.1561/0600000107}. Neural data compression is a machine learning technique that performs the compression task using deep neural networks (DNNs). Current research in neural compression is largely driven by the development of deep generative models such as generative adversarial networks (GANs) \cite{NIPS2014_5ca3e9b1}, variational autoencoders (VAEs) \cite{Kingma2014}, normalizing flows \cite{pmlr-v15-larochelle11a}, and autoregressive models \cite{pmlr-v48-oord16} for image and video compression.
These models have proven effective in capturing complex data distributions, which is crucial for achieving high compression rates while maintaining data fidelity.

In the wireless communication domain, the first neural CSI compressor was introduced in\cite{8322184}, where the authors used a simple convolutional neural network (CNN) auto-encoder architecture for dimension reduction of a massive MIMO channel matrix. Later, numerous network architectures and machine learning techniques have been proposed to further improve the CSI compression efficiency \cite{8972904, 9149229, 9797871, 9705497}. While most works formulated the problem in terms of the dimensions of the reduced CSI matrix, the feedback channels must ultimately convey the CSI matrix in bits, necessitating additional quantization in the latent space.
In \cite{9296555} and \cite{8918798}, the authors formulated the problem within the rate-distortion (RD) framework and further enhanced the performance of the neural CSI compressor by incorporating learned entropy encoding and decoding blocks. They employed a weighted RD loss function to jointly minimize the mean squared error (MSE) and the average bit-rate overhead, explicitly quantifying the communication cost of compression in bits per CSI dimension under a variable-rate compression scheme.

Machine learning models often perform remarkably well on test sets with a distribution similar to their training data but can fail catastrophically in deployment when the data distribution shifts. Such shifts can arise from various factors, including temporal changes, domain variations, or sampling biases, and pose significant challenges in many real-world applications \cite{quinonero2022dataset}.
Several approaches, such as transfer learning \cite{zhuang2020comprehensive}, data augmentation \cite{NEURIPS2018_1d94108e}, and domain adaptation \cite{ganin2015unsupervised}, can be employed to enhance the robustness of machine learning models against distribution shift and improve their generalization capabilities. In transfer learning, models trained on generic datasets are fine-tuned for the target domain. Data augmentation improves generalization by artificially generating diverse training samples. Domain adaptation, on the other hand, aligns source and target domain distributions through techniques such as instance re-weighting or feature alignment.

Distribution shift is a common phenomenon in wireless networks. The statistics of the underlying wireless channel change due to various factors such as user mobility, environmental changes, or variations in interference from other devices. When the channel distribution deviates from initial assumptions or models, the efficiency of algorithms for tasks such as channel estimation \cite{10608175}, signal detection \cite{9103314}, etc can degrade. Consequently, maintaining robust communication requires adaptive strategies that can adjust to these shifts.
In the context of CSI compression for FDD massive MIMO systems, most models in the literature are trained and tested using data from the same environment, e.g., within a specific macro-cell coverage area. However, this would require users to either store or download \cite{9834372} new models for every new cell they enter, making these approaches infeasible in practice. Therefore, it is crucial to develop techniques that can overcome the distribution shift problem in CSI compression.

Several works have studied the effect of distribution shift in the CSI compression problem \cite{9442844, 10508320, 10262359, 10097872, 10381825, 10359472, 9737435}. In \cite{9442844}, a deep transfer learning method is used to handle the distribution shift and a model-agnostic meta-learning algorithm is proposed to reduce the number of CSI samples. In \cite{10508320}, a lightweight translation model and data augmentation method based on domain knowledge are introduced. Specifically, to adapt to new channel conditions, the authors propose an efficient scenario-adaptive CSI feedback architecture, “CSI-TransNet,” and an efficient deep unfolding-based CSI compression network, “SPTM2-ISTANet+.” A single-encoder-to-multiple-decoders (S-to-M) approach is presented in \cite{10262359}, where the authors use multi-task learning to integrate multiple independent autoencoders into a unified architecture featuring a shared encoder and several task-specific decoders.

The authors in \cite{10097872} proposed online learning schemes based on a vanilla autoencoder architecture, exploring various adaptation strategies. In \cite{10381825}, elastic weight consolidation is incorporated to alleviate catastrophic forgetting in NNs. A federated edge learning (FEEL)-based training framework for massive MIMO CSI feedback is introduced in \cite{10359472} to address privacy concerns arising from transmitting raw CSI datasets during centralized training, and a personalization strategy is further developed to enhance feedback performance. In \cite{9737435}, a gossip learning strategy is adopted, and two data augmentation techniques, random erasing and random phase shift, are employed to improve generalization capability.

However, we highlight that the communication overhead associated with model updates is neglected in all the prior works mentioned above. This represents a significant gap in the literature, as typical DNNs can contain millions of parameters, and transmitting model updates can result in massive feedback overhead. In \cite{10359472}, low-resolution quantization (e.g., 2-bit) is applied to the model updates; however, the resulting update bit-rate is not accounted for in the evaluation. To elaborate, the decoder network in \cite{10359472} includes approximately 4 million trainable parameters. Even with an aggressive 2-bit quantization, the resulting communication overhead would be around 8 Mbits, substantially higher than that is required to transmit the compressed CSI matrix.
More precisely, a compressed CSI matrix of size 64$\times$64, using a modest compression ratio of 16, would only require 4 kbits and 16 kbits for 2-bit and 32-bit quantization, respectively. Comparing the communication cost for compressed CSI and model updates illustrates that transmitting model updates can introduce a prohibitive feedback overhead for FDD massive MIMO systems.

To bridge this gap, in this paper, we explicitly account for the communication overhead of model update transmission. We investigate how to efficiently transmit both the model updates and the latent compressed CSI to obtain the best overall RD trade-off. The key differences and contributions of this work are summarized as follows:

\begin{itemize}
    \item Prior works addressing distribution shifts in neural CSI compression primarily focus on vanilla autoencoder-based schemes and overlook bit-level CSI compression as well as the communication cost of transmitting model updates to the decoder \cite{9442844, 10508320, 10262359, 10097872, 10381825, 10359472, 9737435}. In contrast, by incorporating lossless entropy coding for both the compressed CSI latents and the model updates, we explicitly quantify the RD trade-off in neural CSI compression while accounting for the total communication cost, including both CSI feedback and model-update signaling.
    \item To enable efficient entropy coding of model updates, we model the update distribution using a spike-and-slab prior, formulated as a mixture of narrow (spike) and wide (slab) Gaussian components. By assigning a higher probability mass to the spike component, the proposed prior promotes sparse and selective parameter updates, allowing only the most impactful weights to be modified. In addition, we incorporate the bit rate of model updates as a regularization term in the RD loss. Simulation results demonstrate that the combined use of spike-and-slab modeling and update-rate regularization is essential for significantly reducing model-update communication cost while improving overall RD performance.
    \item We evaluate the proposed framework across diverse CSI datasets, including QuaDriGa \cite{6758357}, 3GPP clustered delay line (CDL) \cite{3gpp.38.901}, and DeepMIMO \cite{Alkhateeb2019}, covering both site-specific and stochastic channel models and a wide range of wireless environments. We systematically investigate the impact of the amortized model-update rate over the evaluation horizon, as well as the effects of target-domain dataset size and model-update quantization. Our results reveal a fundamental trade-off between the amortized model-update rate and distortion in dynamic wireless environments, and demonstrate that robust RD performance can be achieved using a few hundred CSI samples, with finer model-update quantization further improving performance.
\end{itemize}

The remainder of this paper is organized as follows. Section~\ref{section2} introduces the backbone neural CSI compressor for an FDD massive MIMO system and evaluates its performance in the absence of distribution shifts. Section~\ref{section3} describes the CSI datasets considered and the proposed neural model fine-tuning framework. Section~\ref{section4} presents numerical results in terms of the RD trade-off, along with a complexity analysis. Finally, Section~\ref{section5} concludes the paper.

Throughout this paper, the following notation is adopted. Bold uppercase letters denote matrices, and bold lowercase letters denote vectors. The transpose operation is represented by the superscript $(\cdot)^T$. The expectation operator is denoted by $\mathbb{E}$, and $\|\cdot\|$ represents the norm. The symbols $\lfloor \cdot \rceil$, $\lfloor \cdot \rfloor$, and $\lceil \cdot \rceil$ denote the rounding, floor, and ceiling operations, respectively.

\section{FDD massive MIMO and Backbone Neural CSI Compressor}\label{section2}
\subsection{FDD Massive MIMO}
We consider an FDD transmission scenario in which a massive MIMO BS equipped with \( N_t \) antennas serves multiple single-antenna users. Downlink transmission is performed using orthogonal frequency division multiplexing (OFDM) over \( N_c \) subcarriers.
For each subcarrier \( m \in \{1,\dots,N_c\} \), let \( \mathbf{h}_m \in \mathbb{C}^{N_t} \) denote the downlink channel vector between the BS and a given user, \( \mathbf{w}_m \in \mathbb{C}^{N_t} \) the corresponding transmit precoding vector, \( x_m \in \mathbb{C} \) the transmitted data symbol, and \( \mathbf{n}_m  \in \mathbb{C} \) the additive noise. The received signal at the user on subcarrier \( m \) is given by
\begin{equation}
    \mathbf{y}_m = \mathbf{h}_m^{T} \mathbf{w}_m x_m + \mathbf{n}_m .
\end{equation}

The downlink CSI across all subcarriers in the spatial--frequency domain is represented by the matrix
\begin{equation}
    \mathbf{H} = \big[ \mathbf{h}_1, \ldots, \mathbf{h}_{N_c} \big] \in \mathbb{C}^{N_t \times N_c}.
\end{equation}

In massive MIMO systems, the channel matrix is high-dimensional, making full CSI feedback prohibitively expensive in terms of uplink bandwidth. To alleviate this bottleneck, CSI compression is commonly employed, whereby the user compresses the CSI matrix and the BS subsequently reconstructs the channel matrix. The objective of CSI compression in FDD massive MIMO systems is to minimize the reconstruction distortion while satisfying a prescribed feedback bit-rate constraint from the user to the BS.

\subsection{Backbone Neural CSI Compressor}
We consider an autoencoder-based NN architecture that explicitly targets the RD trade-off in CSI compression. Quantization, entropy encoding, and entropy decoding are incorporated into the training procedure, enabling optimization of both the feedback rate and the reconstruction distortion of the neural CSI compressor. The proposed compressor is denoted by
\begin{equation}
c = (f_\phi, g_\theta, \gamma_\theta),
\end{equation}
where $f_\phi: \mathcal{C}^{N_t \times N_c} \rightarrow \mathcal{Z}$ represents the feature encoder parameterized by $\phi$, $g_\theta: \mathcal{Z} \rightarrow \mathcal{C}^{N_t \times N_c}$ denotes the feature decoder, and $\gamma_\theta$ corresponds to the entropy coder, jointly parameterized by $\theta$~\cite{10.1561/0600000107}. For notational simplicity, the same symbol $\theta$ is used to denote the parameter set of both the feature decoder and the entropy coder.

The \emph{encoder} maps each CSI matrix $\mathbf{H} \in \mathbb{C}^{N_t \times N_c}$ to a latent representation $\mathbf{Z} \in \mathcal{Z}$. The latent representation is then quantized and transmitted losslessly to the receiver using a variable-length entropy code. Let $\bar{\mathbf{Z}} = Q(\mathbf{Z})$ denote the quantized latent representation, where $Q(\cdot)$ represents the quantization operation. The entropy encoder converts the quantized latent values into a bitstream using a lossless coding scheme, given by
\begin{equation}
b_z = \gamma_\theta(\bar{\mathbf{Z}}; p_\theta),
\end{equation}
where $p_\theta$ denotes the prior probability model of the latent representation.

At the \emph{decoder}, the entropy decoder is first applied to recover the quantized latent representation from the received bitstream as $\bar{\mathbf{Z}} = \gamma_\theta^{-1}(b_z; p_\theta)$. The de-quantization operation is then applied to obtain the continuous latent representation, denoted by $\hat{\mathbf{Z}} = Q^{-1}(\bar{\mathbf{Z}})$. We emphasize that $Q^{-1}(\cdot)$ does not represent a true inverse of the quantization function $Q(\cdot)$ and is generally non-bijective in lossy compression. Both the encoder and decoder employ a shared entropy coder $\gamma_\theta$ with prior distribution $p_\theta$ to ensure the lossless transmission of the quantized latent representation $\bar{\mathbf{Z}}$. Finally, the feature decoder reconstructs the CSI matrix from the de-quantized latent representation as
\begin{equation}
\hat{\mathbf{H}} = g_\theta\!\left(\hat{\mathbf{Z}}\right).
\end{equation}

Let the distortion function be denoted by $\rho: \mathbf{H} \times \hat{\mathbf{H}} \rightarrow [0,\infty)$, which quantifies the reconstruction error between the ground-truth CSI matrix and its estimate. In practice, the distortion is typically measured using the squared error, i.e.,
$\rho(\mathbf{H}, \hat{\mathbf{H}}) \propto \lVert \mathbf{H} - \hat{\mathbf{H}} \rVert^2$.
Under a lossy compression framework, the resulting distortion of a compressor $c$ is defined as
\begin{equation}
D(c) = \mathbb{E}\!\left[\rho(\mathbf{H}, \hat{\mathbf{H}})\right].
\end{equation}

The expectation is taken over the random realization of the channel matrix $\mathbf{H}$, and without loss of optimality, we restrict our attention to deterministic codes \cite{DBLP:journals/corr/abs-2102-09270}. Let the bit length associated with encoding a CSI matrix $\mathbf{H}$ be defined as
\begin{equation}
l(\mathbf{H}) := \bigl| \gamma_\theta\!\left(Q\!\left(f_\phi(\mathbf{H})\right)\right) \bigr|,
\end{equation}
 and the corresponding average bit rate be defined as
\begin{equation}\label{rate_calc}
R(c) = \mathbb{E}\!\left[l(\mathbf{H})\right],
\end{equation}
which captures the expected number of bits required to encode the CSI matrix $\mathbf{H}$. In practice, this rate can be approximated by the entropy of the quantized latent representation, resulting in
\begin{equation}\label{rate_calc_entropy}
R(c) = \mathbb{E}\!\left[-\log_2 p_\theta(\bar{\mathbf{Z}})\right].
\end{equation}

The objective is to jointly minimize the distortion and the rate with respect to the encoder, decoder, and latent representations. According to RD theory, the fundamental limits of lossy compression are characterized by a conditional distribution $p_{\hat{\mathbf{H}}|\mathbf{H}}$. The RD function is given by
\begin{equation}
R(D) = \inf_{p_{\hat{\mathbf{H}}|\mathbf{H}}:\,\mathbb{E}\!\left[\rho(\mathbf{H},\hat{\mathbf{H}})\right]\leq D}
I\!\left(\mathbf{H};\hat{\mathbf{H}}\right),
\end{equation}
where $I(\mathbf{H};\hat{\mathbf{H}})$ denotes the mutual information between the CSI matrix $\mathbf{H}$ and its reconstruction $\hat{\mathbf{H}}$. In theory, this optimal rate can be achieved using vector quantization by jointly compressing many independent realizations of the channel matrix. However, such schemes become intractable for high-dimensional CSI and are incompatible with practical systems, where each CSI matrix must be compressed individually. To address this limitation, we adopt an \emph{operational RD} formulation and denote the set of admissible codecs by $\mathcal{C}$:

\begin{equation}
R_{\mathrm{O}}(D)
= \inf_{c \in \mathcal{C}:\,\mathbb{E}\!\left[\rho(\mathbf{H}, \hat{\mathbf{H}})\right] \leq D}
\mathbb{E}\!\left[l(\mathbf{H})\right],
\end{equation}
which characterizes the \emph{operational RD function} over the class of admissible codecs $\mathcal{C}$. This constrained optimization problem can be equivalently addressed via a Lagrangian formulation,
\begin{equation}\label{Lagrangian_loss}
L(\lambda, c)
= R(c) + \lambda D(c)
= \mathbb{E}\!\left[l(\mathbf{H})\right]
+ \lambda \mathbb{E}\!\left[\rho(\mathbf{H}, \hat{\mathbf{H}})\right].
\end{equation}

For any $\lambda \geq 0$, minimizing the Lagrangian objective over $\mathcal{C}$ yields an optimal codec $c^*$. The parameter $\lambda$ governs the trade-off between rate and distortion. Larger values of $\lambda$ prioritize distortion minimization, resulting in operation at a higher bit-rate regime, whereas smaller values favor stronger compression at the expense of reconstruction accuracy. By adopting the MSE distortion measure, the resulting RD loss is expressed as
\begin{equation}\label{RD_loss_final}
\begin{aligned}
L_{\mathrm{RD}}(\phi,\theta)
&= \mathbb{E}\!\left[
-\log_2 p_\theta\!\left(Q\!\left(f_\phi(\mathbf{H})\right)\right)
\right] \\
&\quad + \lambda\, \mathbb{E}\!\left[
\left\|
g_\theta\!\left(Q^{-1}\!\left(Q(f_\phi(\mathbf{H}))\right)\right)
- \mathbf{H}
\right\|^2
\right].
\end{aligned}
\end{equation}

\begin{figure}
    \centering
    \includegraphics[height=6cm, width=8.5cm]{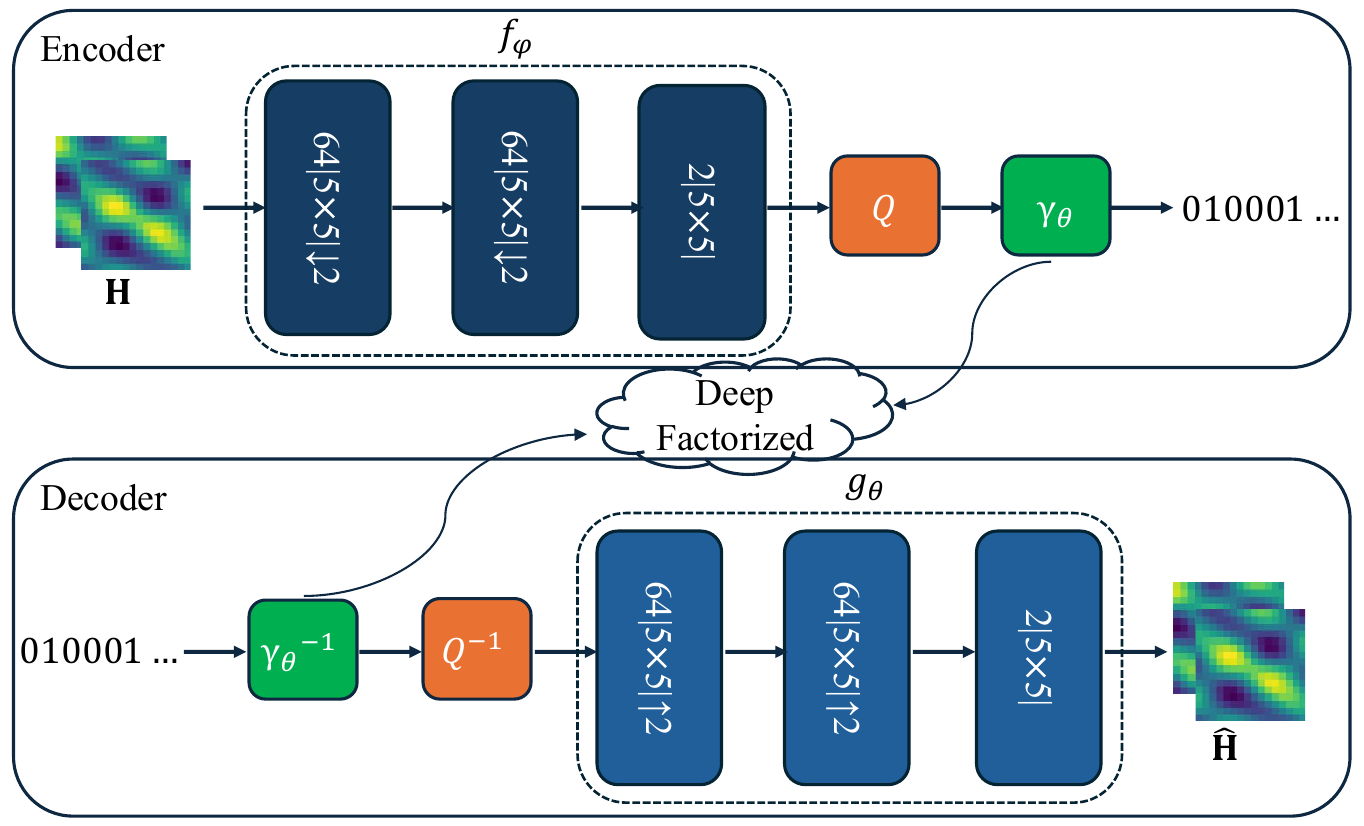}
    \caption{The neural CSI compressor architecture.}
    \label{fig:architecture}
\end{figure}

Fig.~\ref{fig:architecture} illustrates the encoder–decoder architecture of the proposed neural CSI compressor, where the input consists of the real and imaginary components of the spatial–frequency channel matrix. The encoder function $f_\phi$ is composed of a sequence of convolutional layers with nonlinear activation functions that map the high-dimensional massive MIMO CSI matrix $\mathbf{H}$ to a lower-dimensional latent representation. Specifically, the first two layers employ convolutional filters with 64 kernels followed by ReLU activation functions, while the final layer uses 2 convolutional kernels with linear activation. Max-pooling operations are applied to progressively reduce the spatial dimensions of the input CSI tensor.

After obtaining the low-dimensional latent representation, quantization and entropy coding are applied to convert the latent space into a bit sequence. Specifically, a uniform scalar quantizer with unit step size is employed to map continuous latent values to discrete symbols. Leveraging the latent probability distribution learned during training, a lossless entropy coding scheme, such as range coding or context-adaptive binary arithmetic coding (CABAC), is then used to encode the quantized latent representation, resulting in a variable-length bitstream.

Incorporating quantization into the compression framework introduces a challenge for gradient-based optimization, as the quantization operation has a zero derivative almost everywhere. To address this issue, a common relaxation is adopted in which quantization is approximated by adding independent and identically distributed uniform noise, $\Delta \mathbf{Z} \sim \mathcal{U}(0,1)$, to the latent representation, resulting in $\tilde{\mathbf{Z}} = \mathbf{Z} + \Delta \mathbf{Z}$. This relaxation provides a differentiable surrogate and yields an upper bound on the expected number of bits required to encode the latent space. During inference, the true quantization operation is applied, and the rate is computed using the actual quantized latent values without additive noise.

On the decoder side, the entropy-coded bitstream is first decoded to recover the quantized latent representation, which is then mapped to a low-dimensional channel representation by the decoding network. The entropy encoder and decoder share the same latent prior distribution. In our architecture, we adopt a fully factorized prior implemented via the \texttt{DeepFactorized} model \cite{ballé2018variational}, which leverages deep neural networks to learn the latent probability distribution under factorization assumptions.

The decoder network, represented by the function $g_\theta$ in Fig.~\ref{fig:architecture}, consists of a stack of convolutional layers with a kernel size of $5 \times 5$, each followed by a ReLU activation function. To reconstruct the high-dimensional CSI matrix from the low-dimensional entropy-decoded representation, an upsampling operation is applied to progressively restore the spatial resolution.

\section{Neural CSI Compression Fine-Tuning}\label{section3}

A fundamental challenge in learning-based methods is handling distribution shifts. When model capacity is limited or training data is insufficient, such shifts can cause significant performance degradation, thereby limiting real-world effectiveness. This issue is particularly critical in applications such as neural CSI compression for wireless communication systems, where the operating environment is inherently dynamic.
\subsection{Evaluation of The Backbone Neural CSI Compressor}
To assess the performance of the trained neural CSI compressor, we employ the DeepMIMO dataset, which constructs MIMO channel realizations using ray-tracing simulations generated by Remcom Wireless InSite \cite{Remcom}. Specifically, the neural CSI compressor is trained using the static outdoor scenario referred to as {O1} at a carrier frequency of 28~GHz. The detailed parameter configuration for this dataset is summarized in Table~\ref{tab:deepmimo}.

A total of 11{,}920 CSI realizations are generated and partitioned into training, validation, and test sets using a 40\%, 40\%, and 20\% split, respectively. Training is conducted using the RD loss function defined in~(\ref{RD_loss_final}), with the objective of jointly optimizing the encoder and decoder parameters as well as the factorized probability model used for entropy coding. The backbone neural CSI compressor is implemented in Python~3.9 using the TensorFlow framework. Network parameters are optimized using the Adam optimizer with a learning rate of $10^{-3}$ and a batch size of 32 over 200 training epochs.

We assess the performance of the neural CSI compressor in terms of the RD trade-off achieved, where the distortion is measured as the normalized mean squared error (NMSE), defined as 
\begin{equation}
    \text{NMSE} \triangleq \mathbb{E}\left[ \frac{\norm{\mathbf{H} - \mathbf{\hat{H}}}^{2}}{\norm{\mathbf{H}}^{2}}\right],
\end{equation}
where $\mathbf{H}$ and $\hat{\mathbf{H}}$ denote the true and reconstructed channel matrices, respectively. The bit rate of the entropy-coded latent representation is estimated using~(\ref{rate_calc_entropy}) and normalized by the CSI dimensionality, i.e., $64 \times 64$.

The RD curves in Fig.~\ref{fig:backbone_rd} illustrate the performance of the trained neural CSI compressor for different values of $\lambda$, where each operating point corresponds to $\lambda \in {5 \times 10^4, 10^5, 5 \times 10^5, 10^6}$. The results confirm that the proposed neural CSI compressor achieves strong RD performance when evaluated on previously unseen CSI realizations drawn from the same propagation environment.

\begin{table}[t]
    \centering
    \caption{DeepMIMO dataset parameters}
    \begin{tabular}{ll}
        \toprule
        \textbf{DeepMIMO parameters} & \textbf{Value} \\
        \midrule
        Scenario & {O1}\\
        Center frequency &  28 GHz \\
        Number of paths & 10 \\
        Active users & from row 1100 to 2200 \\
        Active BS number & BS 5 \\
        Bandwidth & 50 MHz \\
        Number of OFDM subcarriers & 64 \\
        BS antennas & $N_x$ = 1, $N_y$ = 64, $N_z$ = 1 \\
        UE antennas & $N_x$ = 1, $N_y$ = 1, $N_z$ = 1 \\
        \bottomrule
    \end{tabular}
    \label{tab:deepmimo}
\end{table}


\begin{figure}[t]
\centering
\begin{tikzpicture}

\pgfplotstableread[
  col sep=comma,
]{Results/baseline_rd.csv}\baselineRD

\begin{axis}[
    width=.48\textwidth,
    height=0.35\textwidth,
    xlabel={Rate (bits)},
    ylabel={NMSE (dB)},
    grid=both,
    grid style={dashed,gray!25},
    tick label style={font=\footnotesize},
    label style={font=\small},
    every axis plot/.append style={line width=0.9pt},
    every mark/.append style={mark size=2.2pt},
    scaled x ticks=false,
    xtick distance=0.1,
    ytick distance=4,
    xticklabel style={
      /pgf/number format/fixed,
      /pgf/number format/precision=3
    },
]
\addplot[
  very thick,
  mark=*,
  draw=blue, mark options={draw=blue, fill=blue},
]
table[
  x index=0,
  y index=1,
  skip first n=1,
]{\baselineRD};

\end{axis}
\end{tikzpicture}
\caption{RD performance of the backbone neural CSI compressor.}
\label{fig:backbone_rd}
\end{figure}
\subsection{Evaluation Under Distribution Shifts}
To analyze the impact of distribution shifts, we evaluate the trained neural CSI compressor in propagation environments that differ significantly from the training domain. Specifically, we consider multiple distinct datasets, namely QuaDRiGa~\cite{6758357}, 3GPP CDL~\cite{3gpp.38.901}, and DeepMIMO~\cite{Alkhateeb2019} (under a different scenario), to assess the resulting RD performance. In the following, we describe the key parameter configurations and channel characteristics of these datasets.

\begin{figure}[t]
    \centering
    \includegraphics[height=5cm,width=8cm]{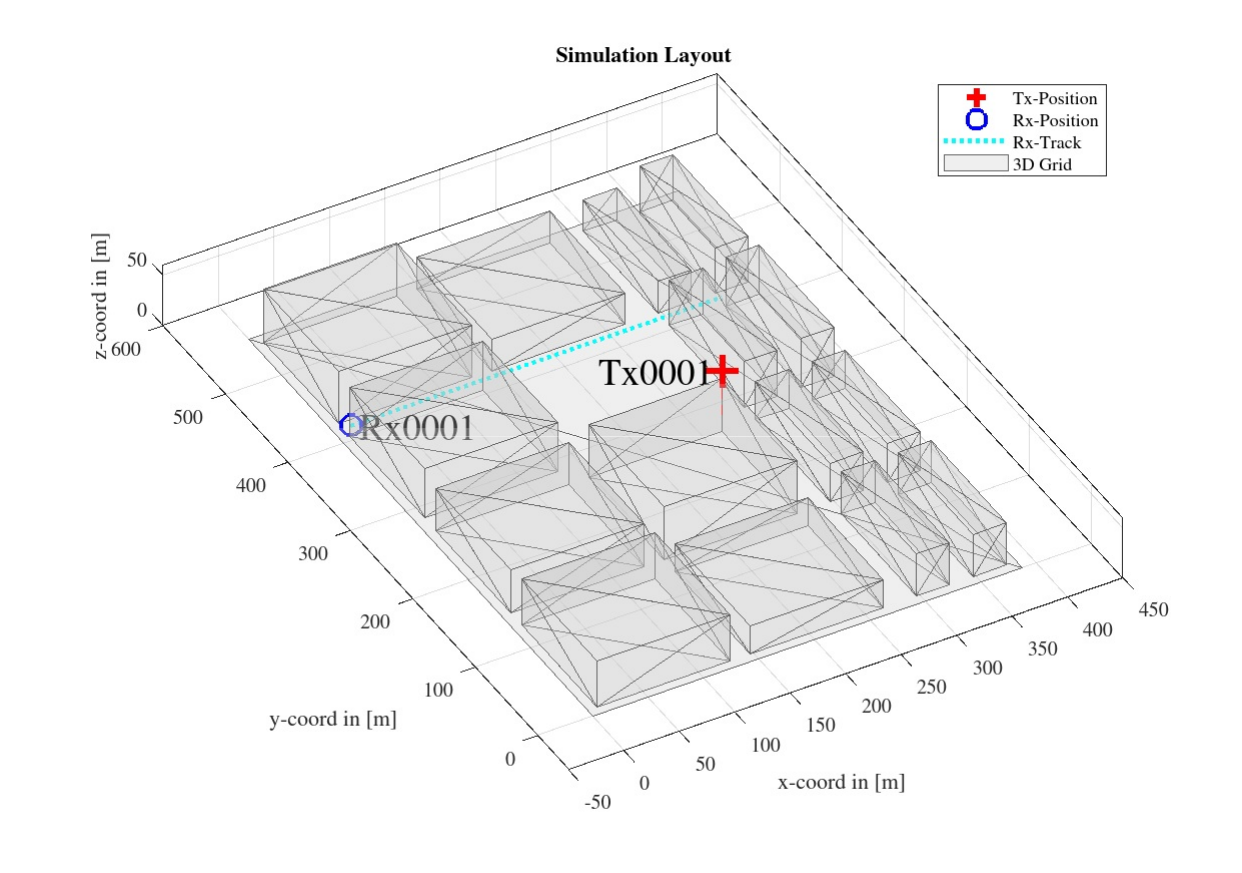}
    \caption{Simulation layout of QuaDRiGa dataset \cite{6758357}.}
    \label{fig:Sim_layout_QuaDRiGa}
\end{figure}

\paragraph{QuaDRiGa Dataset}
QuaDRiGa is a geometry-based stochastic spatial channel model that captures spatial consistency and realistic multipath evolution in three-dimensional propagation environments. Using the QuaDRiGa channel generator, we simulate an urban massive MIMO deployment with a BS equipped with 64 antenna elements and operating over 64 OFDM subcarriers. The BS is placed on a rooftop, while a mobile user traverses a linear track of length 350~m. The underlying three-dimensional propagation environment is based on the Madrid grid developed within the METIS project, enabling realistic modeling of spatial channel evolution and path dynamics along the user trajectory. The simulation layout is illustrated in Fig.~\ref{fig:Sim_layout_QuaDRiGa}, and the corresponding dataset parameters are summarized in Table~\ref{tab:QuaDRiGa}.

\begin{table}[t]
    \centering
    \caption{Simulation parameters of the QuaDRiGa dataset.}
    \label{tab:QuaDRiGa}
    \begin{tabular}{l l}
        \toprule
        \textbf{Parameter} & \textbf{Value} \\
        \midrule
        Carrier frequency & 3.7~GHz \\
        Bandwidth & 50~MHz \\
        Number of OFDM subcarriers & 64 \\
        Number of BS antennas & 64 \\
        Number of UE antennas & 1 \\
        Initial user position & (15, 415, 1.2) \\
        BS position & (267, 267, 60) \\
        BS orientation & $\pi/2$ (facing north) \\
        Spatial sampling density & 29.6~samples/m \\
        \bottomrule
    \end{tabular}
\end{table}

\paragraph{3GPP CDL Dataset}
To further evaluate the performance of the proposed neural CSI compression framework under controlled and standardized channel conditions, we generate a dataset based on the 3GPP CDL channel model defined in TR~38.901. Specifically, we consider the CDL-B delay profile under an urban macrocell (UMa) scenario with a user velocity of 30~km/h. The CDL model characterizes time-varying multipath propagation through a finite set of clusters with fixed angular statistics, while temporal channel evolution is primarily governed by Doppler effects induced by user mobility.
In our setup, the BS is equipped with 64 transmit antennas, and the user employs a single receive antenna. The system operates at a carrier frequency of 28~GHz using an OFDM numerology with a subcarrier spacing of 30~kHz. The detailed simulation parameters for this dataset are summarized in Table~\ref{tab:CDL}.

\begin{table}[t]
\centering
\caption{Simulation parameters for the 3GPP CDL dataset.}
\label{tab:CDL}
\begin{tabular}{l c}
\hline
\textbf{Parameter} & \textbf{Value} \\
\hline
Delay profile & CDL-B \\
Carrier frequency & 28~GHz \\
User velocity & 30~km/h \\
Delay spread & 200~ns \\
Number of BS antennas & 64 \\
Number of UE antennas & 1 \\
OFDM subcarrier spacing & 30~kHz \\
Number of resource blocks & 25 \\
Number of selected subcarriers & 64 \\
CSI sample stride & 5 OFDM symbols \\
Number of CSI samples & 5{,}000 \\
\hline
\end{tabular}
\end{table}

\begin{figure}[t]
    \centering
    \includegraphics[height=4.5cm, width=8.5cm]{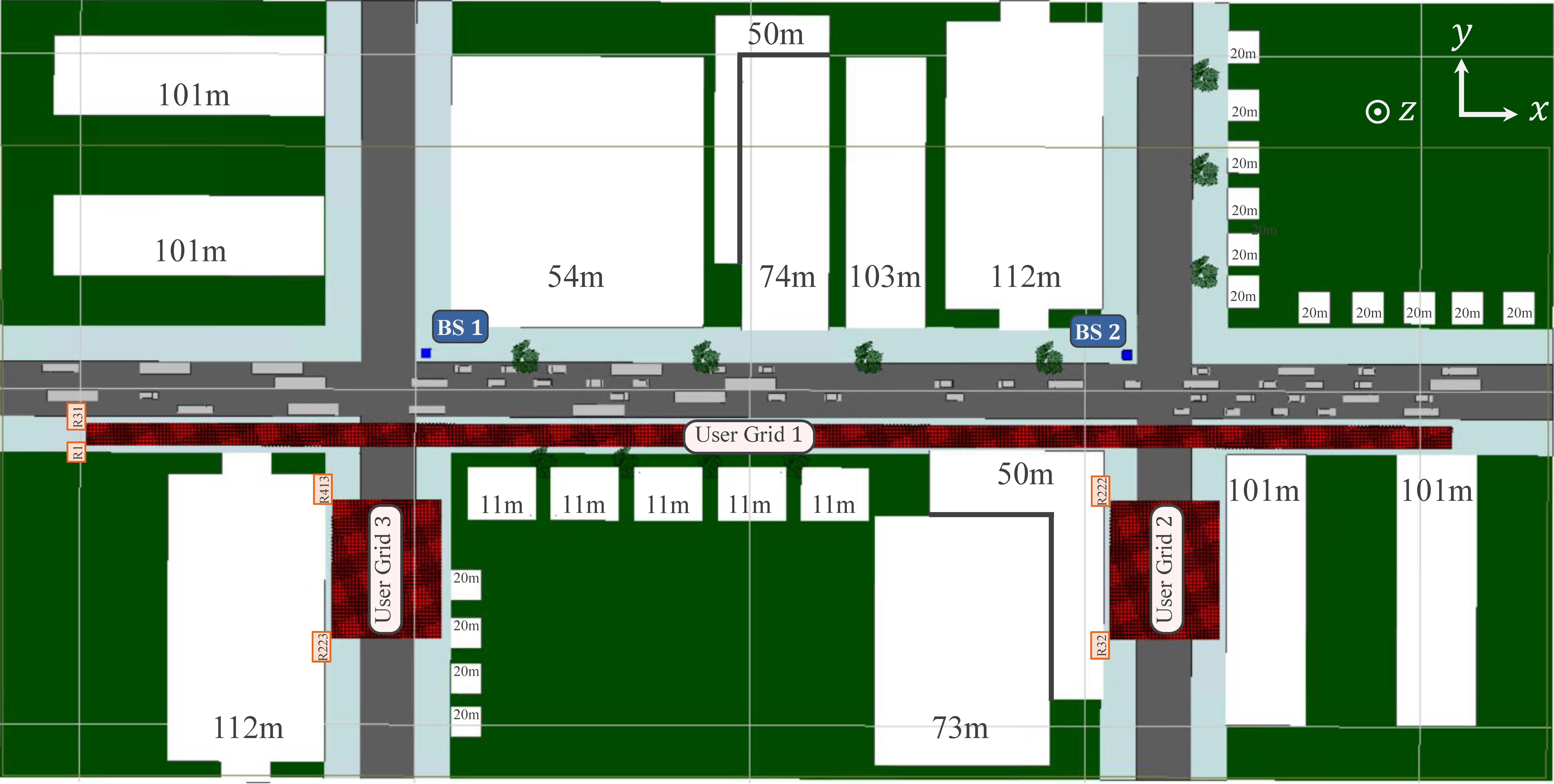}
    \caption{The top view of the {O2 Dynamic} scenario \cite{Alkhateeb2019}.}
    \label{fig:O2_topV3}
\end{figure}

\paragraph{DeepMIMO Dataset}
For the DeepMIMO dataset, we consider another site-specific, ray-tracing-based channel model referred to as {O2 Dynamic}. In this setting, CSI realizations are generated across a sequence of discrete scenes with a sampling interval of 100~ms. The top view of the scenario, shown in Fig.~\ref{fig:O2_topV3}, consists of three streets and two intersections in an urban environment. Across the 1{,}000 captured scenes, vehicular users change positions while the underlying propagation geometry remains fixed, resulting in quasi-stationary channel statistics within each scene. The dataset includes two BSs and approximately 115{,}000 candidate user locations, enabling large-scale and diverse CSI generation under realistic site-specific conditions. The detailed simulation parameters for this scenario are summarized in Table~\ref{tab:deepmimo_dynamic}.

\begin{table}[t]
    \centering
    \caption{Simulation parameters of the DeepMIMO dataset ({O2 Dynamic} scenario).}
    \label{tab:deepmimo_dynamic}
    \begin{tabular}{l l}
        \toprule
        \textbf{Parameter} & \textbf{Value} \\
        \midrule
        Scenario & {O2 Dynamic}  \\
        Carrier frequency & 3.5~GHz \\
        Bandwidth & 50~MHz \\
        Number of OFDM subcarriers & 64 \\
        Number of multipath components & 10 \\
        Active user rows & 1--31 \\
        Scene index & 1 \\
        Active BS & BS~1 \\
        BS antenna configuration & $N_x=1,\; N_y=64,\; N_z=1$ \\
        UE antenna configuration & $N_x=1,\; N_y=1,\; N_z=1$ \\
        \bottomrule
    \end{tabular}
\end{table}

To evaluate the generalization capability of the trained neural CSI compressor in previously unseen environments, we assess its RD performance in Fig.~\ref{fig:rd_noft}. The model is tested on the aforementioned datasets without any fine-tuning. As shown in the figure, the trained compressor exhibits a severe performance degradation under distribution shifts, failing to cope with the new channel statistics. These results clearly demonstrate the limitations of direct model reuse and underscore the necessity of an effective model fine-tuning strategy for neural CSI compression in massive MIMO systems.

\begin{figure}[t]
\centering
\begin{tikzpicture}
\begin{axis}[
  width=.48\textwidth,
  height=0.35\textwidth,
  xlabel={Rate (bits)},
  ylabel={NMSE (dB)},
  grid=both,
  grid style={dashed,gray!25},
  tick label style={font=\footnotesize},
  label style={font=\small},
  scaled x ticks=false,
  unbounded coords=discard,
  legend style={at={(0.5,1.10)}, anchor=south, draw=none, fill=none, font=\footnotesize},
  legend columns=3,
]

\addplot[very thick, mark=*, draw=blue, mark options={draw=blue, fill=blue}]
  table[col sep=comma, x index=0, y index=1, skip first n=1]{Results/NoFT_Quadriga.csv};
\addlegendentry{QuaDRiGa}

\addplot[very thick, mark=square*, draw=red, mark options={draw=red, fill=red}]
  table[col sep=comma, x index=0, y index=1, skip first n=1]{Results/NoFT_CDL.csv};
\addlegendentry{CDL}

\addplot[very thick, mark=triangle*, draw=green!60!black, mark options={draw=green!60!black, fill=green!60!black}]
  table[col sep=comma, x index=0, y index=1, skip first n=1]{Results/NoFT_DeepMIMO.csv};
\addlegendentry{DeepMIMO}

\end{axis}
\end{tikzpicture}
\caption{Evaluation of the backbone neural CSI compressor using QuaDRiGa, CDL, and DeepMIMO datasets.}
\label{fig:rd_noft}
\end{figure}

\subsection{Neural Model Fine-Tuning}
A trained model can be fine-tuned to new channel statistics by exposing it to data samples from the target environment. In the context of neural CSI compression, model fine-tuning corresponds to updating the backbone network parameters, namely \(\phi_0\) and \(\theta_0\) for the encoder and decoder, respectively, which were initially trained on a generic dataset. Fine-tuning the backbone neural CSI compressor using CSI samples from a new domain yields updated encoder and decoder parameters, denoted by \(\phi\) and \(\theta\), respectively. Since the encoder is assumed to be fixed at the transmitter, only the updated decoder parameters need to be conveyed in the bitstream.

We consider a full-model fine-tuning, where both the encoder and decoder networks are fine-tuned while explicitly accounting for the additional communication overhead associated with conveying decoder model updates from the transmitter to the receiver. Specifically, the decoder model updates, defined as $\delta \triangleq \theta - \theta_{0}$, are encoded jointly with the latent representation $z$.
To encode the decoder model update vector, it is first quantized. Unlike the unit-bin quantization applied to the latent space, a higher-resolution quantization is employed for the model updates, since their values typically vary only slightly, depending on the learning rate. In particular, $N$ equispaced bins of width $t$ are used, and the quantization function for the model updates is defined as follows \cite{rozendaal2021overfitting}:
\begin{equation}
    \Bar{\delta} = Q_t(\delta) = \text{clip}\left( \left\lfloor \frac{\delta}{t} \right\rceil t, 
-\frac{(N-1)t}{2}, \frac{(N-1)t}{2} \right),
\end{equation}
where clipping and rounding operations are defined as:
\begin{equation}
\text{clip}(x, x_{\text{min}}, x_{\text{max}}) =
\begin{cases} 
x_{\text{min}}, & \text{if } x < x_{\text{min}}, \\
x, & \text{if } x_{\text{min}} \leq x \leq x_{\text{max}}, \\
x_{\text{max}}, & \text{if } x > x_{\text{max}},
\end{cases}
\end{equation}
\begin{equation}
\lfloor x \rceil =
\begin{cases} 
\lfloor x \rfloor, & \text{if } x - \lfloor x \rfloor < 0.5, \\
\lceil x \rceil, & \text{if } x - \lfloor x \rfloor \geq 0.5.
\end{cases}
\end{equation}

Since both rounding and clipping are non-differentiable operations, they obstruct gradient-based training of the neural CSI compressor. To address this issue, a commonly used technique known as the straight-through estimator (STE) is employed \cite{DBLP:journals/corr/BengioLC13}, where the gradient of the quantization function is approximated as
\(\partial Q_t(\delta) / \partial \delta = 1\).
Here, the quantization bin width is controlled by \(t\), and \(N\) denotes the number of quantization bins.

Let \(\bar{\delta} = Q_t(\delta)\) denote the quantized model update. The discrete model prior \(p[\bar{\delta}]\) is obtained by pushing forward the continuous prior \(p(\delta)\) through the quantization function \(Q_t\), yielding
\begin{equation}
\begin{split}
p[\bar{\delta}] &= \int_{Q_t^{-1}(\bar{\delta})} p(\delta) \, d\delta \\
&= \int_{\bar{\delta} - t/2}^{\bar{\delta} + t/2} p(\delta) \, d\delta \\
&= P(\delta < \bar{\delta} + t/2) - P(\delta < \bar{\delta} - t/2).
\end{split}
\end{equation}

Thus, \(p[\bar{\delta}]\) represents the probability mass assigned to the quantization bin centered at \(\bar{\delta}\). It is computed as the difference between the cumulative distribution function (CDF) values of \(p(\delta)\) evaluated at the bin boundaries.

After quantization, an appropriate model prior \( p(\delta) \) must be selected to enable efficient entropy coding of the discrete model updates. Various distributions can be used to model this prior, such as a zero-mean Gaussian, i.e., \( p(\delta) = \mathcal{N}(\mathbf{0}, \sigma \mathbf{I}) \). However, a key limitation of a Gaussian prior is its relatively high coding cost for zero updates. As a result, even when no effective update is applied, encoding such updates can still incur a non-negligible bit rate.

To address this inefficiency, a spike-and-slab prior is adopted \cite{Ročková02012018}, defined as
\begin{equation}
p(\delta) = \frac{p_{\mathrm{slab}}(\delta) + \alpha\, p_{\mathrm{spike}}(\delta)}{1 + \alpha},
\end{equation}
where
\begin{equation}
p_{\mathrm{slab}}(\delta) = \mathcal{N}(\delta \mid 0, \sigma^2 \bm{I})
\end{equation}
denotes the slab component, and
\begin{equation}
p_{\mathrm{spike}}(\delta) = \mathcal{N}(\delta \mid 0, (t/6)^2 \bm{I})
\end{equation}
denotes the spike component.

The resulting prior \( p(\delta) \) is a mixture of Gaussian distributions, where \( \alpha \in \mathbb{R}^{+} \) controls the relative weight of the spike component. The parameters \( t \) and \( \sigma \), with \( \sigma \gg t/6 \), correspond to the standard deviations of the spike and slab distributions, respectively. By setting the standard deviation of the spike component to \( t/6 \), approximately \(99.7\%\) of its probability mass lies within the central quantization bin after quantization. Consequently, selecting a sufficiently large value of \( \alpha \) significantly reduces the bit-rate cost associated with zero updates and encourages the model to learn only the most informative parameter updates. This prior effectively promotes sparsity in the model updates while maintaining flexibility for larger, informative deviations.

\begin{algorithm}[t]
\caption{Full-Model Fine-Tuning — Encoding}
\label{alg:FM_EN}
\textbf{Input:} Global model parameters $\{\theta_0, \phi_0\}$ trained on a generic dataset; 
batch size $B$; CSI samples from a new environment $\mathcal{H}=\{\mathcal{H}_{\mathrm{T}}, \mathcal{H}_{\mathrm{E}}\}$. \\
\textbf{Output:} Compressed bitstream $\mathbf{b}=(b_{\bar{\delta}}, b_z)$.

\begin{algorithmic}[1]
    \STATE Initialize $\phi \gets \phi_0$, $\theta \gets \theta_0$
    \FOR{i $= 1$ to epochs}
        \FOR{each mini-batch $\mathcal{B} \subset \mathcal{H}_{\mathrm{T}}$ with $|\mathcal{B}|=B$}
            \STATE Load data $\mathbf{H} \in \mathcal{B}$
            \STATE Compute $\delta \gets \theta - \theta_0$, quantize $\bar{\delta} \gets Q_t(\delta)$, and set $\bar{\theta} \gets \theta_0 + \bar{\delta}$
            \STATE Encode features $\mathbf{Z} \gets f_{\phi}(\mathbf{H})$ and quantize $\tilde{\mathbf{Z}} \gets \mathbf{Z} + \Delta\mathbf{Z}$
            \STATE Decode $\hat{\mathbf{H}} \gets g_{\theta}(\tilde{\mathbf{Z}})$
            \STATE Compute loss $L_{\mathrm{RDM}}(\phi,\theta)$
            \STATE Backpropagate with STE and update $\phi, \theta$ using gradients $\nabla_{\phi,\theta} L_{\mathrm{RDM}}(\phi,\theta)$
        \ENDFOR
    \ENDFOR
    \STATE Fine-tuned parameters: $\phi^\ast \gets \phi$, $\theta^\ast \gets \theta$
    \STATE For evaluation set $\mathcal{H}_{\mathrm{E}}$: compute $\bar{\mathbf{Z}} \gets Q(f_{\phi^\ast}(\mathbf{H}))$
    \STATE Quantize tuned parameters: $\bar{\delta} \gets Q_t(\theta^\ast - \theta_0)$, $\bar{\theta} \gets \theta_0 + \bar{\delta}$
    \STATE Entropy encode: $b_{\bar{\delta}} \gets \gamma(\bar{\delta};p(\bar{\delta}))$, \ $b_z \gets \gamma_{\bar{\theta}}(\bar{\mathbf{Z}};p_{\bar{\theta}})$
    \STATE \textbf{Return} $\mathbf{b}=(b_{\bar{\delta}}, b_z)$
\end{algorithmic}
\end{algorithm}

\begin{algorithm}
\caption{Full-Model Fine-Tuning — Decoding}
\label{alg:FM_DE}
\textbf{Input:} Global parameters $\theta_0$ trained on a generic dataset; 
model prior $p(\bar{\delta})$; 
bitstream $\mathbf{b}=(b_{\bar{\delta}}, b_z)$. \\
\textbf{Output:} Decompressed CSI matrix $\hat{\mathbf{H}}$.

\begin{algorithmic}[1]
    \STATE Entropy decode $\bar{\delta} \gets \gamma^{-1}(b_{\bar{\delta}};p(\bar{\delta}))$
    \STATE Compute updated decoder parameters $\bar{\theta} \gets \theta_0 + \bar{\delta}$
    \STATE Entropy decode latent $\bar{\mathbf{Z}} \gets \gamma_{\bar{\theta}}^{-1}(b_z;p_{\bar{\theta}})$
    \STATE De-quantize and decode: $\hat{\mathbf{H}} \gets Q^{-1}(g_{\bar{\theta}}(\bar{\mathbf{Z}}))$
    \STATE \textbf{Return} $\hat{\mathbf{H}}$
\end{algorithmic}
\end{algorithm}

The bit-rate cost associated with the quantized model updates $\bar{\delta}$, under the discrete prior $p[\bar{\delta}]$, is given by
\begin{equation}
    \bar{M} = -\log_2 p[\bar{\delta}].
\end{equation}
Following common practice, this discrete cost is approximated during training by its continuous counterpart \cite{rozendaal2021overfitting},
\begin{equation}
    M = -\log_2 p(\delta).
\end{equation}

To regularize the bit-rate cost of model updates during full-model fine-tuning, the update cost is incorporated into the RD objective in \eqref{RD_loss_final} as
\begin{equation}
L_{\mathrm{RDM}}(\phi,\theta)
= L_{\mathrm{RD}}(\phi,\theta)
- \log_2 p(\delta),
\end{equation}

By optimizing this objective during full-model fine-tuning, the network learns to balance the achievable RD gain against the bit-rate cost required to transmit the decoder model updates. Model fine-tuning is performed using CSI samples collected in the target environment, denoted by $\mathcal{H}_{\mathrm{T}}$. For evaluation, the fine-tuned model is applied to CSI samples acquired after the fine-tuning phase, denoted by $\mathcal{H}_{\mathrm{E}}$.
The encoding and decoding procedures for full-model fine-tuning are summarized in Algorithms~\ref{alg:FM_EN} and~\ref{alg:FM_DE}, respectively.

\section{Performance Evaluation}\label{section4}

In this section, we evaluate the effectiveness of different model-fine-tuning schemes applied to the backbone neural CSI compressor in target environments with shifted channel statistics, along with the associated computational cost of model fine-tuning. We consider the following neural CSI compression fine-tuning schemes:

\begin{itemize}
    \item \textbf{Pretrained}: denotes applying the pretrained neural CSI compressor directly in the target environment without any model fine-tuning.

    \item \textbf{EO}: corresponds to the encoder-only fine-tuning scheme. Encoder parameters are updated using CSI samples from the target environment, while the decoder remains fixed. This enables partial specialization of the pretrained model to the target environment without requiring any additional feedback. Results are reported for models trained at different values of $\lambda$. 

    \item \textbf{FM}: refers to the full-model fine-tuning scheme, where both encoder and decoder parameters are updated. The spike-and-slab prior parameters are set to $t=0.005$, $N=50$, and $\sigma=0.05$. Only the low-bit-rate neural CSI compressor trained at $\lambda = 5 \times 10^{4}$ is fine-tuned. The reported rate for this scheme represents the total rate, i.e., the sum of the latent bit-rate and the amortized model-update bit-rate. Minor parameter adjustments are made to achieve the best performance.

    \item \textbf{FM1}: Motivated by the observation that earlier layers of DNNs capture more general features, whereas later layers encode finer, task-specific details \cite{10018052}, model fine-tuning is restricted to the final convolutional layer of the decoder network to reduce the bit-rate cost of model updates. Accordingly, {FM1} denotes a partial full-model fine-tuning scheme in which only the final convolutional layer of the decoder is updated, while all other layers remain fixed. The same update parameters are used as in the {FM} scheme.

    \item \textbf{FM1-UP}: is identical to {FM1}, except that a uniform prior is used for the model updates instead of the spike-and-slab prior, as originally proposed in our initial work \cite{11143260}. Since only a single decoder layer is updated, resulting in a smaller number of model update parameters compared to {FM}, the uniform prior imposes weaker sparsity constraints and may therefore better accommodate a larger number of effective updates.

    \item \textbf{FM-2bit}: represents the full-model fine-tuning scheme with coarse quantization using four quantization bins (2-bit quantization). Prior results in \cite{10359472}, based on a federated edge learning framework, suggest that the impact of coarse quantization noise can be mitigated in the downlink stream. Motivated by this observation, the FM scheme is evaluated under this coarse quantization setting.

    \item \textbf{GA}: denotes a genie-aided scheme, in which the decoder is assumed to have perfect knowledge of the model updates without any feedback overhead. This scheme provides a lower bound on the achievable performance of neural CSI compression under distribution shifts. Similarly to {FM}, only the low-bit-rate trained neural CSI compressor is used. The approaches in \cite{9442844,10097872} can be interpreted as instances of this genie-aided setting, as they update a vanilla autoencoder without accounting for the communication overhead of model-update feedback.

    \item \textbf{TM}: corresponds to the translation-module-based scheme proposed in \cite{10508320}. Inspired by image-to-image translation in computer vision, this approach maps input CSI to the target domain using a convolutional translation module at the encoder and a retranslation module at the decoder. The same network architecture and sparsity-alignment function as in \cite{10508320} are employed. Similarly to the encoder-only fine-tuning, models trained at different values of $\lambda$ are used. To provide a lower-bound rate, the communication overhead required to convey the retranslation module parameters to the decoder is also ignored.
\end{itemize}
\subsection{RD Numerical Results}
In this subsection, we present the RD results of the different fine-tuning schemes for neural CSI compression. The schemes are evaluated using the QuaDRiGa, 3GPP CDL, and DeepMIMO datasets introduced in Section~\ref{section3}. Fine-tuning is performed using 100 CSI samples from the target domain with a batch size of 50 over 100 epochs.

\definecolor{cNoFT}{RGB}{0,114,178}
\definecolor{cEO}{RGB}{213,94,0}
\definecolor{cFMten}{RGB}{0,158,115}
\definecolor{cFMLast}{RGB}{204,121,167}
\definecolor{cFMLastUP}{RGB}{230,159,0}
\definecolor{cGA}{RGB}{86,180,233}
\definecolor{cFM}{RGB}{150,60,180}
\definecolor{cTM}{RGB}{200,30,60}

\begin{figure*}[t]
\centering

\begin{tikzpicture}
\begin{groupplot}[
  group style={group size=3 by 1, horizontal sep=8mm},
  width=0.35\textwidth,
  height=0.30\textwidth,
  xlabel={Rate (bits)},
  ylabel={NMSE (dB)},
  grid=both,
  grid style={dashed,gray!25},
  tick label style={font=\footnotesize},
  label style={font=\small},
  title style={font=\small},
  every axis plot/.append style={line width=0.9pt},
  every mark/.append style={mark size=2.2pt},
  scaled x ticks=false,
  xtick distance=0.1,
  ytick distance=4,
  xticklabel style={
    /pgf/number format/fixed,
    /pgf/number format/precision=3
  },
  legend columns=4,
  legend style={
    /tikz/every even column/.append style={column sep=3pt},
    draw=none, fill=none, font=\small
  },
]

\newcommand{\plotmethod}[5]{%
  \addplot[
    color=#2,
    mark=#3,
    discard if not={method}{#1}
  ] table[
    col sep=comma,
    x=rate,
    y=nmse_db
  ] {#5};
  \addlegendentry{#4}
}

\nextgroupplot[title={QuaDRiGa}, legend to name=RDLegend]

\plotmethod{NoFT}{cNoFT}{o}{Pretrained}{Results/QuaDRiGa.csv}
\plotmethod{EO}{cEO}{x}{EO}{Results/QuaDRiGa.csv}
\plotmethod{TM}{cTM}{*}{TM}{Results/QuaDRiGa.csv}
\plotmethod{FM}{cFM}{triangle*}{FM}{Results/QuaDRiGa.csv}
\plotmethod{FM10bits}{cFMten}{diamond*}{FM-2bit}{Results/QuaDRiGa.csv}
\plotmethod{FMLastLayer}{cFMLast}{pentagon*}{FM1}{Results/QuaDRiGa.csv}
\plotmethod{FMLastLayerUP}{cFMLastUP}{star}{FM1-UP}{Results/QuaDRiGa.csv}
\plotmethod{GA}{cGA}{square*}{GA}{Results/QuaDRiGa.csv}

\nextgroupplot[title={3GPP CDL}, ylabel={}]

\addplot[color=cNoFT, mark=o, discard if not={method}{NoFT}]
  table[col sep=comma, x=rate, y=nmse_db] {Results/CDL_30.csv};
\addplot[color=cEO, mark=x, discard if not={method}{EO}]
  table[col sep=comma, x=rate, y=nmse_db] {Results/CDL_30.csv};
\addplot[color=cTM, mark=*, discard if not={method}{TM}]
  table[col sep=comma, x=rate, y=nmse_db] {Results/CDL_30.csv};
\addplot[color=cFM, mark=triangle*, discard if not={method}{FM}]
  table[col sep=comma, x=rate, y=nmse_db] {Results/CDL_30.csv};
\addplot[color=cFMten, mark=diamond*, discard if not={method}{FM10bits}]
  table[col sep=comma, x=rate, y=nmse_db] {Results/CDL_30.csv};
\addplot[color=cFMLast, mark=pentagon*, discard if not={method}{FMOneLayer}]
  table[col sep=comma, x=rate, y=nmse_db] {Results/CDL_30.csv};
\addplot[color=cFMLastUP, mark=star, discard if not={method}{FMOneLayerUP}]
  table[col sep=comma, x=rate, y=nmse_db] {Results/CDL_30.csv};
\addplot[color=cGA, mark=square*, discard if not={method}{GA}]
  table[col sep=comma, x=rate, y=nmse_db] {Results/CDL_30.csv};

\nextgroupplot[title={DeepMIMO}, ylabel={}]

\addplot[color=cNoFT, mark=o, discard if not={method}{NoFT}]
  table[col sep=comma, x=rate, y=nmse_db] {Results/DeepMIMO.csv};
\addplot[color=cEO, mark=x, discard if not={method}{EO}]
  table[col sep=comma, x=rate, y=nmse_db] {Results/DeepMIMO.csv};
\addplot[color=cTM, mark=*, discard if not={method}{TM}]
  table[col sep=comma, x=rate, y=nmse_db] {Results/DeepMIMO.csv};
\addplot[color=cFM, mark=triangle*, discard if not={method}{FM}]
  table[col sep=comma, x=rate, y=nmse_db] {Results/DeepMIMO.csv};
\addplot[color=cFMten, mark=diamond*, discard if not={method}{FM10bits}]
  table[col sep=comma, x=rate, y=nmse_db] {Results/DeepMIMO.csv};
\addplot[color=cFMLast, mark=pentagon*, discard if not={method}{FMOneLayer}]
  table[col sep=comma, x=rate, y=nmse_db] {Results/DeepMIMO.csv};
\addplot[color=cFMLastUP, mark=star, discard if not={method}{FMOneLayerUP}]
  table[col sep=comma, x=rate, y=nmse_db] {Results/DeepMIMO.csv};
\addplot[color=cGA, mark=square*, discard if not={method}{GA}]
  table[col sep=comma, x=rate, y=nmse_db] {Results/DeepMIMO.csv};

\end{groupplot}

\node[anchor=south, yshift=16pt]
  at ($(group c1r1.north)!0.5!(group c3r1.north)$)
  {\pgfplotslegendfromname{RDLegend}};

\end{tikzpicture}

\caption{RD results of different fine-tuning schemes evaluated on multiple datasets.}

\label{fig:rd-3datasets}

\end{figure*}

Fig.~\ref{fig:rd-3datasets} compares the RD performance of different fine-tuning schemes across the QuaDRiGa, 3GPP CDL, and DeepMIMO datasets.
From the figure, it can be observed that the gains achieved by EO remain limited compared to schemes that permit decoder-side updates. This confirms that encoder-only fine-tuning can only partially compensate for distribution shifts, while the decoder remains mismatched to the target environment, thereby highlighting the importance of full-model fine-tuning. Moreover, the TM scheme, which relies on translation and retranslation modules without fine-tuning the backbone network, does not provide noticeable RD gains, and its performance remains largely comparable to that of the pretrained model.
On the other hand, full-model fine-tuning yields substantial RD improvements across all datasets, demonstrating the benefit of jointly updating the encoder and decoder parameters. Notably, the reported rates for FM include both the latent bit-rate and the amortized model-update bit-rate, indicating that these improvements are achieved despite explicitly accounting for the communication overhead associated with decoder updates.

The partial full-model fine-tuning scheme (FM1 and FM1-UP), which restricts updating to the final convolutional layer of the decoder, underperforms full FM but outperforms EO and TM. This confirms that fine-tuning only high-level, task-specific layers is insufficient to fully capture environment-specific channel characteristics compared to full-model fine-tuning; nevertheless, it still provides an RD improvement relative to schemes in which the decoder remains unchanged.
Furthermore, a comparison between FM1 and FM1-UP highlights that when only a single decoder layer is updated, weaker sparsity constraints on the model updates can be afforded, and the resulting RD performance is comparable to that achieved by FM1 with the spike-and-slab prior.

A comparison between FM and FM-2bits suggests that employing finer quantization for model updates improves RD performance. This observation contrasts with the results reported in \cite{10359472} and can primarily be attributed to differences in the system setup. In particular, \cite{10359472} considers a multi-user scenario with a federated learning framework, where quantization noise introduced by individual users can be partially compensated through aggregation, whereas such compensation is not present in our considered single-model fine-tuning setting.

Finally, the genie-aided scheme (GA) consistently provides the best RD performance and serves as a lower bound, as it assumes perfect decoder-side knowledge of model updates without any feedback overhead. The performance gap between GA and FM quantifies the cost of explicitly accounting for model-update transmission and underscores the efficiency of the proposed RDM-based full-model fine-tuning framework.

%
%
%
\definecolor{cQuadLine}{RGB}{0,114,178}   
\definecolor{cRMaLine}{RGB}{213,94,0}     
\definecolor{cDeepLine}{RGB}{0,158,115}   

\definecolor{cAnnoA}{RGB}{0,0,0}          
\definecolor{cAnnoB}{RGB}{90,90,90}       
\definecolor{cAnnoC}{RGB}{120,60,0}       

\begin{figure*}[t]
\centering

\begin{tikzpicture}
\begin{groupplot}[
  group style={group size=3 by 1, horizontal sep=8mm},
  width=0.35\textwidth,
  height=0.30\textwidth,
  xlabel={Rate (bits)},
  ylabel={NMSE (dB)},
  grid=both,
  grid style={dashed,gray!25},
  tick label style={font=\footnotesize},
  label style={font=\small},
  title style={font=\small},
  every axis plot/.append style={line width=0.9pt},
  every mark/.append style={mark size=2.2pt},
  scaled x ticks=false,
  xticklabel style={
    /pgf/number format/fixed,
    /pgf/number format/precision=3
  },
  unbounded coords=discard,
  clip=false,
]

\nextgroupplot[title={QuaDRiGa}]

\addplot+[
  thick,
  color=cQuadLine,
  mark=*,
  mark size=2.2pt,
  nodes near coords={\scriptsize $\pgfplotspointmeta\,\mathrm{m}$},
  point meta=explicit symbolic,
  every node near coord/.append style={
    text=cAnnoA,
    anchor=west,
    yshift=-7pt,
    xshift={
      \ifnum\pgfplotspointmeta=16
        -12pt
      \else
        -5pt
      \fi
    }
  },
] table[
  col sep=comma,
  x index=1,
  y index=2,
  meta index=0
] {Results/QuaDRiGa_meters.csv};

\nextgroupplot[title={3GPP CDL}, ylabel={}]

\addplot+[
  thick,
  color=cRMaLine,
  mark=square*,
  mark size=2.2pt,
  nodes near coords={\scriptsize $\pgfplotspointmeta$ ms},
  point meta=explicit symbolic,
  every node near coord/.append style={
    text=cAnnoA,
    anchor=west,
    xshift=-15pt,
    yshift=7pt
  },
] table[
  col sep=comma,
  x index=1,
  y index=2,
  meta index=0
] {Results/RMa_120_time_ms.csv};

\nextgroupplot[title={DeepMIMO}, ylabel={}]

\addplot+[
  thick,
  color=cDeepLine,
  mark=triangle*,
  mark size=2.4pt,
  nodes near coords={\scriptsize $\pgfplotspointmeta\,\mathrm{s}$},
  point meta=explicit symbolic,
  every node near coord/.append style={
    text=cAnnoA,
    anchor=west,
    xshift=0pt,
    yshift=1pt
  },
] table[
  col sep=comma,
  x expr={ (\thisrowno{0}==50) ? nan : \thisrowno{1} },
  y expr={ (\thisrowno{0}==50) ? nan : \thisrowno{2} },
  meta index=0
] {Results/DeepMIMO_scene.csv};

\end{groupplot}
\end{tikzpicture}

\caption{RD trade-off of full-model fine-tuning across different evaluation horizons. Point annotations indicate distance (m) for QuaDRiGa, channel evolution time at 120~km/h for the CDL model, and captured scene duration for DeepMIMO.}

\label{fig:rd-evolution}
\end{figure*}

Fig.~\ref{fig:rd-evolution} illustrates the RD performance across evolving CSI samples in the QuaDRiGa, 3GPP CDL, and DeepMIMO datasets. Each operating point corresponds to a different evaluation horizon, defined as the duration over which an updated model is reused before the CSI distribution changes sufficiently to require further updates. The total feedback rate accounts for both the compressed CSI and the amortized communication cost of model updates over the evaluation horizon.

In the QuaDRiGa dataset, channel evolution is driven by user mobility along a 350~m linear track in a realistic three-dimensional urban environment, where propagation paths may appear and disappear as the user moves. This leads to pronounced structural changes in the CSI distribution over distance. Consequently, increasing the evaluation horizon results in a noticeable degradation in reconstruction quality unless model updates are transmitted more frequently. This behavior manifests as a clear trade-off between the amortized model-update rate and distortion, highlighting the limited validity period of fine-tuned models in highly nonstationary propagation environments.

In contrast, the 3GPP CDL dataset exhibits a smooth temporal channel evolution dominated by Doppler effects, with a fixed propagation geometry and no abrupt path changes. As a result, the updated model remains effective over longer time spans, and the RD performance improves as the amortized model-update rate decreases with longer evaluation horizons. It is worth noting that, in this experiment, the user velocity is increased to 120~km/h and a rural macro-cell (RMa) delay profile is employed in order to accelerate CSI variations and create a more challenging evaluation scenario.
Similarly, in the DeepMIMO O2 Dynamic scenario, CSI realizations are collected across discrete scenes in a site-specific urban layout with fixed geometry within each scene. Although the user environment varies across scenes, the underlying propagation characteristics remain largely unchanged, leading to quasi-stationary CSI statistics. As a result, the fine-tuned model can be reused over long evaluation horizons with limited degradation in distortion performance.


\pgfplotsset{
  discard if not/.style 2 args={
    x filter/.code={
      \edef\temp{\thisrow{#1}}%
      \edef\val{#2}%
      \ifx\temp\val
      \else
        \def\pgfmathresult{nan}%
      \fi
    }
  }
}

\definecolor{cSSP}{RGB}{0,114,178}
\definecolor{cWO}{RGB}{0,158,115}
\definecolor{cUP}{RGB}{200,30,60}
\definecolor{cTot}{RGB}{150,60,180}
\definecolor{cUpd}{RGB}{86,180,233}
\begin{figure*}[t]
\centering
\begin{tikzpicture}

\begin{groupplot}[
  group style={group size=3 by 1, horizontal sep=12mm},
  width=0.35\textwidth,
  height=0.30\textwidth,
  grid=both,
  grid style={dashed,gray!25},
  scaled x ticks=false,
  tick label style={font=\footnotesize},
  label style={font=\small},
  title style={font=\small},
  xlabel style={font=\small, yshift=-1pt},
  ylabel style={font=\small, yshift=0pt},
  y label style={at={(axis description cs:-0.12,0.5)}, anchor=south}
]

\nextgroupplot[
  xlabel={Rate (bits)},
  ylabel={NMSE (dB)},
  every axis plot/.append style={line width=0.7pt},
  every mark/.append style={mark size=2.3pt},
  legend style={
    at={(0.5,1.1)}, anchor=south,
    draw=none, fill=none,
    font=\footnotesize,
    legend columns=2,
    /tikz/every even column/.append style={column sep=6pt},
    row sep=1pt
  }
]

\newcommand{\plotrd}[4]{%
  \addplot[
    color=#2,
    mark=#3,
    discard if not={Method}{#1}
  ] table[
    col sep=comma,
    x=Rate,
    y=NMSEdB
  ] {Results/rd_ablation.csv};
  \addlegendentry{#4}
}

\plotrd{FM_L_RDM}{cSSP}{o}{FM-SSP}
\plotrd{FM_L_RD}{cWO}{square*}{FM w/o reg.}
\plotrd{FM_UP}{cUP}{triangle*}{FM-UP}

\nextgroupplot[
  xlabel={Epoch},
  ylabel={Non-zero updates},
  xmin=0,
  xmax=200,
  ymode=log,
  log basis y=10,
  every axis plot/.append style={line width=1pt},
  legend style={
    at={(0.5,1.1)}, anchor=south,
    draw=none, fill=none,
    font=\footnotesize,
    legend columns=2,
    /tikz/every even column/.append style={column sep=8pt},
    row sep=1pt
  }
]

\addplot[color=cSSP] table[col sep=comma, x=Step, y=NonZeroUpdates_SSP] {Results/non_zero_updates.csv};
\addlegendentry{SSP}
\addplot[color=cUP]  table[col sep=comma, x=Step, y=NonZeroUpdates_UP]  {Results/non_zero_updates.csv};
\addlegendentry{UP}

\nextgroupplot[
  xlabel={Epoch},
  ylabel={Rate (bits)},
  xmin=0,
  xmax=200,
  every axis plot/.append style={line width=1pt},
  legend style={
    at={(0.5,1.1)}, anchor=south,
    draw=none, fill=none,
    font=\footnotesize,
    legend columns=2,
    /tikz/every even column/.append style={column sep=8pt},
    row sep=1pt
  }
]

\addplot[color=cTot] table[col sep=comma, x=Step, y expr=\thisrow{TotalRate_SSP}/4096.0] {Results/rates_model_updates_total.csv};
\addlegendentry{T-SSP}
\addplot[color=cWO]  table[col sep=comma, x=Step, y expr=\thisrow{TotalRate_UP}/4096.0]  {Results/rates_model_updates_total.csv};
\addlegendentry{T-UP}
\addplot[dashed] table[col sep=comma, x=Step, y expr=\thisrow{ModelUpdates_SSP}/4096.0] {Results/rates_model_updates_total.csv};
\addlegendentry{M-SSP}
\addplot[color=cUP, dashed]  table[col sep=comma, x=Step, y expr=\thisrow{ModelUpdates_UP}/4096.0]  {Results/rates_model_updates_total.csv};
\addlegendentry{M-UP}

\end{groupplot}
\end{tikzpicture}

\caption{Ablation results for full-model scheme, highlighting the role of the spike-and-slab prior and the model-update regularizer in controlling the model-update rate.}

\label{fig:ablation}

\end{figure*}
Fig.~\ref{fig:ablation} presents an ablation study that examines the impact of the model-update prior and the update-rate regularizer on full-model fine-tuning. The scheme denoted FM-SSP corresponds to full-model fine-tuning with a spike-and-slab prior on the model updates, whereas FM-UP employs a uniform prior. The variant FM w/o reg. removes the model-update regularizer from the RD loss, thereby allowing unrestricted model updates. For clarity, SSP and UP refer to the spike-and-slab and uniform priors, respectively.
To analyze the rate behavior in detail, the total feedback rate is decomposed into its components. The terms T-SSP and T-UP denote the total rate, including both the latent bit-rate and the amortized model-update rate, when using the spike-and-slab and uniform priors, respectively. In contrast, M-SSP and M-UP represent only the model-update rate associated with each prior.

The results in Fig.~\ref{fig:ablation} lead to several conclusions regarding full-model fine-tuning. First, incorporating the model-update regularizer in the RD loss is essential for controlling the bit-rate cost of model updates. When the regularizer is removed (FM w/o reg.), the model-update rate increases rapidly, resulting in a substantially higher total rate without corresponding RD gains. This shows that unconstrained full-model fine-tuning can lead to an excessive increase in the model-update bit rate.

Second, the choice of prior has a pronounced impact on the sparsity and rate of model updates. The spike-and-slab prior consistently yields fewer nonzero updates and a lower model-update rate compared to the uniform prior, while achieving strong RD performance. This confirms that explicitly promoting sparsity in model updates is critical to reducing the feedback overhead associated with model fine-tuning.

\begin{table}[t]
\centering
\renewcommand{\arraystretch}{1.2}
\caption{Impact of model-update quantization resolution on RD performance.}
\begin{tabular}{c|cc|cc}
\hline
\multirow{2}{*}{\textbf{Quantization}} 
& \multicolumn{2}{c|}{\textbf{QuaDRiGa}} 
& \multicolumn{2}{c}{\textbf{3GPP CDL}} \\
& Rate & NMSE (dB) & Rate & NMSE (dB) \\
\hline
1 bit & 0.25 &  -5.73  & 0.232 &  -7.88  \\  
2 bit & 0.34 &  -8.81  & 0.308 & -12.52  \\
3 bit & 0.33 & -12.05  & 0.300 & -16.59  \\
4 bit & 0.32 & -13.70  & 0.287 & -17.56  \\
5 bit & 0.30 & -13.89  & 0.277 & -17.59  \\
6 bit & 0.30 & -13.89  & 0.276 & -17.59  \\
7 bit & 0.30 & -13.89  & 0.276 & -17.59  \\
\hline
\end{tabular}
\label{tab:RD_quantization_row}
\end{table}

Table~\ref{tab:RD_quantization_row} reports the RD performance of the full-model scheme under different quantization resolutions for model updates, evaluated on the QuaDRiGa and CDL datasets. The results indicate that low-resolution quantization (below 3 bits) leads to substantial RD degradation. In contrast, increasing the quantization resolution beyond 4 bits does not yield any noticeable RD improvement, suggesting diminishing returns from high-resolution quantization. In general, moderate-resolution quantization is sufficient to achieve good compression performance.

\begin{table}[t]
\centering
\renewcommand{\arraystretch}{1.2}
\caption{RD performance for different dataset size for full-model fine-tuning.}
\label{tab:dataset_size_results}
\begin{tabular}{c|cc|cc}
\hline
\multirow{2}{*}{\textbf{Dataset Size}} 
& \multicolumn{2}{c|}{\textbf{QuaDRiGa}} 
& \multicolumn{2}{c}{\textbf{3GPP CDL}} \\
& Rate & NMSE (dB) & Rate & NMSE (dB) \\
\hline
1    & 0.248 &  -7.03 & 0.241 & -10.48 \\
10   & 0.286 & -10.38 & 0.267 & -15.09 \\
50   & 0.299 & -11.79 & 0.269 & -15.26 \\
100  & 0.297 & -13.89 & 0.275 & -17.64 \\
200  & 0.305 & -14.63 & 0.288 & -18.86 \\
300  & 0.316 & \textbf{-13.96} & 0.302 & -19.98 \\
500  & 0.325 & \textbf{-10.39} & 0.330 & -20.75 \\
\hline
\end{tabular}
\end{table}

The results in Table~\ref{tab:dataset_size_results} illustrate the impact of the dataset size on the RD performance of the full-model fine-tuning scheme. The results indicate that increasing the number of CSI samples used for fine-tuning initially leads to clear improvements in RD performance. However, beyond a certain dataset size, further increasing the number of training samples yields only marginal performance gains, suggesting that a few hundred CSI samples are sufficient for effective full-model fine-tuning.
In the QuaDRiGa dataset, using a larger number of CSI samples (e.g., 300–500 samples) leads to a degradation in NMSE performance. This behavior is attributed to the increased heterogeneity of the fine-tuning data, as larger training sets include more diverse channel geometries and spatial conditions. As a result, the fine-tuned model becomes less locally specialized, which is suboptimal for the fixed evaluation horizon considered in this study.

\begin{table}[t]
\centering
\caption{Sensitivity analysis of the spike-and-slab prior hyperparameters.}
\label{tab:ss_sensitivity}
\renewcommand{\arraystretch}{1.2}
\setlength{\tabcolsep}{4pt}
\begin{tabular}{c|c|c|c|c}
\hline
$\alpha$ & $\sigma$ & $t$ & Rate & NMSE (dB) \\
\hline

\multicolumn{5}{c}{$t$ sweep ($\alpha$=100, $\sigma$=0.05)} \\
\hline
100 & 0.05 & 0.0005 & 0.577 & -9.60 \\
100 & 0.05 & 0.001           & 0.346 & -12.90 \\
100 & 0.05 & 0.005                & 0.314 & -13.59 \\
100 & 0.05 & 0.01                 & 0.275 & -13.17 \\
100 & 0.05 & 0.05                 & 0.225 & -6.25 \\
\hline

\multicolumn{5}{c}{$\sigma$ sweep ($\alpha$=100, $t$=0.005)} \\
\hline
100 & 0.005 & 0.005 & NaN & NaN \\
100 & 0.01  & 0.005 & inf & -13.71 \\
100 & 0.05  & 0.005 & 0.314 & -13.59 \\
100 & 0.1   & 0.005 & 0.317 & -13.68 \\
100 & 0.5   & 0.005 & 0.320 & -13.53 \\
100 & 1     & 0.005 & 0.319 & -13.58 \\
\hline

\multicolumn{5}{c}{$\alpha$ sweep ($\sigma$=0.005, $t$=0.05)} \\
\hline
0.1   & 0.005 & 0.05 & 0.506 & -13.50 \\
1     & 0.005 & 0.05 & 0.363 & -13.79 \\
10    & 0.005 & 0.05 & 0.315 & -13.73 \\
100   & 0.005 & 0.05 & 0.314 & -13.59 \\
1000  & 0.005 & 0.05 & 0.320 & -13.55 \\
10000 & 0.005 & 0.05 & 0.322 & -13.54 \\
\hline
\end{tabular}
\end{table}

Table~\ref{tab:ss_sensitivity} evaluates the sensitivity of the spike-and-slab prior hyperparameters $(\alpha,\sigma,t)$ on RD performance for QuaDRiGa dataset. 
We vary one hyperparameter at a time while keeping the others fixed, as indicated in each block.
From the $t$-sweep, we observe that extremely small values of $t$ lead to a very strong shrinkage towards zero, which destabilizes the update distribution and results in an increase in the rate and degraded NMSE. In this regime, moderate-magnitude updates are excessively penalized and inefficiently encoded through the slab component. On the other hand, excessively large $t$ weakens the sparsity-inducing effect of the spike component, reducing the separation between near-zero and significant updates and degrading the RD performance. A broad stable region is observed for $t \in [10^{-3}, 10^{-2}]$, where the RD trade-off is well balanced.

For the $\sigma$-sweep, very small slab variances lead to numerical instability, as the slab component becomes excessively narrow and cannot accommodate moderate update magnitudes. This results in unstable entropy estimates and unreliable coding behavior. In contrast, for $\sigma \geq 0.05$, the RD performance remains stable, with only minor variations across more than one order of magnitude, indicating that the method is not sensitive to the precise choice of slab scale within this range.
Finally, $\alpha$ -sweep shows that $\alpha$ primarily governs the balance between the spike and slab components, thereby controlling the effective sparsity level of the updates. 
Across a wide range ($\alpha \in [1,10^4]$), the RD trade-off exhibits only modest variation, with diminishing changes for large $\alpha$.

\subsection{Computational Complexity}
We analyze the computational cost of the proposed full-model fine-tuning by
explicitly accounting for the number of floating-point operations (FLOPs)
required by the neural CSI compressor considered in this work. The training
complexity of model fine-tuning naturally depends on the chosen backbone
architecture, with simpler or more complex networks incurring lower or higher
computational costs, respectively. In this study, the backbone neural CSI
compressor consists of stacked $5\times5$ convolutional and transposed
convolutional layers operating on CSI tensors of spatial resolution
$64\times64$.

For a 2D convolution producing an output tensor of size
$H\times W\times C_{\text{out}}$ with kernel size $k\times k$ and
$C_{\text{in}}$ input channels, the number of multiply--accumulate operations
(MACs) is given by $HWC_{\text{out}}(k^2C_{\text{in}})$, where one MAC is counted
as two FLOPs. Summing the contributions of all
convolutional layers, a single forward pass through the encoder--decoder
architecture requires approximately $0.48$~GFLOPs per CSI sample. To account for
backpropagation and gradient computation during fine-tuning, we adopt a standard
approximation whereby the total training cost is three times the forward-pass
complexity, resulting in approximately $1.43$~GFLOPs per CSI sample per fine-tuning
iteration \cite{kaplan2020scalinglawsneurallanguage}.

The on-device update is performed using $N_T=100$ CSI samples, a batch size of
$B=50$, and $E=100$ epochs, corresponding to two optimization steps per epoch and
$200$ steps in total. Under these settings, the total fine-tuning cost is
given by
\begin{equation}
    \text{FLOPs}_{\text{total}} \approx 1.43 \times N_T \times E
    \approx 14.25~\text{TFLOPs}.
\end{equation}

The expected fine-tuning latency can be approximated as
$T \approx \text{FLOPs}_{\text{total}}/\eta$, where $\eta$ denotes the effective
sustained throughput of the mobile processor in TFLOPs/s~\cite{11106171}. For
typical mobile devices, $\eta$ may range from $0.1$ to $1$~TFLOPs/s depending on
the available hardware acceleration and numerical precision, resulting in
fine-tuning latencies on the order of several tens of seconds.

The corresponding energy consumption scales linearly with the total number of
operations and can be expressed as
$E \approx \text{FLOPs}_{\text{total}}\epsilon_{\text{FLOP}}$, where
$\epsilon_{\text{FLOP}}$ denotes the energy consumed per floating-point
operation. For representative values of $\epsilon_{\text{FLOP}}$ between
$10$ and $100$~pJ/FLOP~\cite{11106171}, the total energy cost of fine-tuning lies in the range of $0.04$--$0.4$~Wh, corresponding to a small fraction
of a typical smartphone battery capacity.
In practical deployments, the fine-tuning will be executed as a background
process and invoked only occasionally, e.g., when significant distribution
shifts are detected. As a result, the fine-tuning cost can be amortized over time
and is unlikely to interfere with latency-critical operations.

\section{Conclusion}\label{section5}
A major limitation of neural compression approaches is their significant performance degradation when channel statistics change, which is a natural phenomenon in wireless environments. To address this distribution shift problem, we proposed a fine-tuning scheme for neural CSI compression that explicitly accounts for the communication overhead associated with transmitting model updates. Lossless entropy coding is applied to both latent representations and model updates, and a spike-and-slab prior is adopted to promote sparse and efficient parameter updates. Furthermore, the bit-rate of model updates is incorporated into the fine-tuning process through a regularized RD loss function.
We conducted simulations across a wide range of wireless CSI datasets, covering both site-specific and stochastic channel models. The results demonstrate that full-model fine-tuning is essential for effectively adapting the neural CSI compressor to varying environments.

We further evaluated full-model fine-tuning under different evaluation horizons, quantization resolutions, and fine-tuning sample sizes. The results show that in environments with rapidly varying CSI statistics, applying fine-tuned models over excessively long horizons can lead to noticeable RD degradation. In addition, using highly diverse CSI samples for fine-tuning may reduce model specialization within a localized operating region. We also observe that moderate quantization resolutions, on the order of 4–5 bits, are sufficient to achieve strong RD performance.
As a future direction, this work can be extended to fully online learning of neural CSI compression. In this setting, the encoder would dynamically adapt its parameters in response to evolving CSI statistics.


\ifCLASSOPTIONcaptionsoff
  \newpage
\fi

\bibliographystyle{IEEEtran}
\bibliography{Mybib}

@ARTICLE{6736761,
  author={Larsson, Erik G. and Edfors, Ove and Tufvesson, Fredrik and Marzetta, Thomas L.},
  journal={IEEE Commun. Mag.}, 
  title={Massive {MIMO} for Next Generation Wireless Systems}, 
  year={2014},
  volume={52},
  number={2},
  pages={186--195},
  doi={10.1109/MCOM.2014.6736761}
}

@ARTICLE{5595728,
  author={Marzetta, Thomas L.},
  journal={IEEE Trans. Wireless Commun.}, 
  title={Noncooperative Cellular Wireless with Unlimited Numbers of Base Station Antennas}, 
  year={2010},
  volume={9},
  number={11},
  pages={3590--3600},
  doi={10.1109/TWC.2010.092810.091092}
}

@ARTICLE{6816089,
  author={Rao, Xiongbin and Lau, Vincent K. N.},
  journal={IEEE Trans. Signal Process.}, 
  title={Distributed Compressive {CSIT} Estimation and Feedback for {FDD} Multi-User Massive {MIMO} Systems}, 
  year={2014},
  volume={62},
  number={12},
  pages={3261--3271},
  doi={10.1109/TSP.2014.2324991}
}

@ARTICLE{10608175,
  author={Sattari, Mehdi and Guo, Hao and Gündüz, Deniz and Panahi, Ashkan and Svensson, Tommy},
  journal={IEEE Trans. Mach. Learn. Commun. Netw.}, 
  title={Full-Duplex Millimeter Wave {MIMO} Channel Estimation: A Neural Network Approach}, 
  year={2024},
  volume={2},
  number={},
  pages={1093-1108},
  doi={10.1109/TMLCN.2024.3432865}
}

@ARTICLE{9103314,
  author={Khani, Mehrdad and Alizadeh, Mohammad and Hoydis, Jakob and Fleming, Phil},
  journal={IEEE Trans. Wirel. Commun.}, 
  title={Adaptive Neural Signal Detection for Massive {MIMO}}, 
  year={2020},
month = {May.},
  volume={19},
  number={8},
  pages={5635-5648},
  doi={10.1109/TWC.2020.2996144}
}

@ARTICLE{10130108,
  author={Hu, Zhengyang and Liu, Guanzhang and Xie, Qi and Xue, Jiang and Meng, Deyu and Gündüz, Deniz},
  journal={IEEE Trans. Wireless Commun.}, 
  title={A Learnable Optimization and Regularization Approach to Massive {MIMO} {CSI} Feedback}, 
  year={2024},
  volume={23},
  number={1},
  pages={104--116},
  doi={10.1109/TWC.2023.3275990}
}

@misc{wu2024mimochannelneuralfunction,
  title={{MIMO} Channel as a Neural Function: Implicit Neural Representations for Extreme {CSI} Compression in Massive {MIMO} Systems}, 
  author={Haotian Wu and Maojun Zhang and Yulin Shao and Krystian Mikolajczyk and Deniz Gündüz},
  year={2024},
  eprint={2403.13615},
  archivePrefix={arXiv},
  primaryClass={cs.IT},
  url={https://arxiv.org/abs/2403.13615}, 
}

@ARTICLE{10208156,
  author={Rizzello, Valentina and Nerini, Matteo and Joham, Michael and Clerckx, Bruno and Utschick, Wolfgang},
  journal={IEEE Wireless Commun. Lett.}, 
  title={User-Driven Adaptive {CSI} Feedback With Ordered Vector Quantization}, 
  year={2023},
  volume={12},
  number={11},
  pages={1956--1960},
  doi={10.1109/LWC.2023.3301992}
}

@ARTICLE{6497019,
  author={Bhagavatula, Ramya and Heath, Robert W.},
  journal={IEEE Trans. Wireless Commun.}, 
  title={Predictive Vector Quantization for Multicell Cooperation With Delayed Limited Feedback}, 
  year={2013},
  volume={12},
  number={6},
  pages={2588--2597},
  doi={10.1109/TWC.2013.040413.112037}
}

@ARTICLE{9296555,
  author={Mashhadi, Mahdi Boloursaz and Yang, Qianqian and Gündüz, Deniz},
  journal={IEEE Trans. Wireless Commun.}, 
  title={Distributed Deep Convolutional Compression for Massive {MIMO} {CSI} Feedback}, 
  year={2021},
  volume={20},
  number={4},
  pages={2621--2633},
  doi={10.1109/TWC.2020.3043502}
}

@ARTICLE{7442899,
  author={Huang, Xin-Lin and Wu, Jun and Wen, Yonggang and Hu, Fei and Wang, Yi and Jiang, Tao},
  journal={IEEE Trans. Wireless Commun.}, 
  title={Rate-Adaptive Feedback With Bayesian Compressive Sensing in Multiuser {MIMO} Beamforming Systems}, 
  year={2016},
  volume={15},
  number={7},
  pages={4839--4851},
  doi={10.1109/TWC.2016.2547861}
}

@ARTICLE{8322184,  
author={C. {Wen} and W. {Shih} and S. {Jin}},  
journal={IEEE Wireless Commun. Lett.},  
title={Deep Learning for Massive {MIMO} {CSI} Feedback},   
year={2018},
month = {Mar.},
volume={7},  
number={5},  
pages={748-751},}

@article{10.1561/0600000107,
author = {Yang, Yibo and Mandt, Stephan and Theis, Lucas},
title = {An Introduction to Neural Data Compression},
year = {2023},
issue_date = {Apr 2023},
publisher = {Now Publishers Inc.},
address = {Hanover, MA, USA},
volume = {15},
number = {2},
issn = {1572-2740},
url = {https://doi.org/10.1561/0600000107},
doi = {10.1561/0600000107},
journal = {Found. Trends. Comput. Graph. Vis.},
month = {apr},
pages = {113–200},
numpages = {100}
}

@inproceedings{NIPS2014_5ca3e9b1,
  author = {Goodfellow, Ian and Pouget-Abadie, Jean and Mirza, Mehdi and Xu, Bing and Warde-Farley, David and Ozair, Sherjil and Courville, Aaron and Bengio, Yoshua},
  booktitle = {Adv. Neural Inf. Process. Syst.},
  editor = {Z. Ghahramani and M. Welling and C. Cortes and N. Lawrence and K.Q. Weinberger},
  pages = {},
  publisher = {Curran Associates, Inc.},
  title = {Generative Adversarial Nets},
  url = {https://proceedings.neurips.cc/paper_files/paper/2014/file/5ca3e9b122f61f8f06494c97b1afccf3-Paper.pdf},
  volume = {27},
  year = {2014}
}

@inproceedings{Kingma2014,
  author = {Kingma, Diederik P. and Welling, Max},
  booktitle = {2nd Int. Conf. Learn. Represent. (ICLR) 2014, Banff, AB, Canada, Apr. 14-16, 2014, Conf. Track Proc.},
  eprint = {http://arxiv.org/abs/1312.6114v10},
  eprintclass = {stat.ML},
  eprinttype = {arXiv},
  title = {{Auto-Encoding Variational Bayes}},
  year = 2014
}

@InProceedings{pmlr-v15-larochelle11a,
  title = {The Neural Autoregressive Distribution Estimator},
  author = {Larochelle, Hugo and Murray, Iain},
  booktitle = {Proc. 14th Int. Conf. Artif. Intell. Stat. (AISTATS)}, 
  pages = {29--37},
  year = {2011},
  editor = {Gordon, Geoffrey and Dunson, David and Dudík, Miroslav},
  volume = {15},
  series = {Proc. Mach. Learn. Res.},
  address = {Fort Lauderdale, FL, USA},
  month = {11--13 Apr},
  publisher = {PMLR},
  url = {https://proceedings.mlr.press/v15/larochelle11a.html}
}

@InProceedings{pmlr-v48-oord16,
  title = {Pixel Recurrent Neural Networks},
  author = {van den Oord, Aäron and Kalchbrenner, Nal and Kavukcuoglu, Koray},
  booktitle = {Proc. 33rd Int. Conf. Mach. Learn. (ICML)}, 
  pages = {1747--1756},
  year = {2016},
  editor = {Balcan, Maria Florina and Weinberger, Kilian Q.},
  volume = {48},
  series = {Proc. Mach. Learn. Res.},
  address = {New York, NY, USA},
  month = {20--22 Jun},
  publisher = {PMLR},
  url = {https://proceedings.mlr.press/v48/oord16.html}
}

@ARTICLE{8972904,
  author={Guo, Jiajia and Wen, Chao-Kai and Jin, Shi and Li, Geoffrey Ye},
  journal={IEEE Trans. Wireless Commun.},
  title={Convolutional Neural Network-Based Multiple-Rate Compressive Sensing for Massive {MIMO} {CSI} Feedback: Design, Simulation, and Analysis},
  year={2020},
  volume={19},
  number={4},
  pages={2827-2840},
  doi={10.1109/TWC.2020.2968430}
}

@inproceedings{9149229,
  author={Lu, Zhilin and Wang, Jintao and Song, Jian},
  booktitle={Proc. IEEE Int. Conf. Commun. (ICC)},
  title={Multi-resolution {CSI} Feedback with Deep Learning in Massive {MIMO} System}, 
  year={2020},
  volume={},
  number={},
  pages={1-6},
  doi={10.1109/ICC40277.2020.9149229}
}

@ARTICLE{9797871,
  author={Tang, Shunpu and Xia, Junjuan and Fan, Lisheng and Lei, Xianfu and Xu, Wei and Nallanathan, Arumugam},
  journal={IEEE Trans. Veh. Technol.}, 
  title={Dilated Convolution Based {CSI} Feedback Compression for Massive {MIMO} Systems}, 
  year={2022},
  volume={71},
  number={10},
  pages={11216-11221},
  doi={10.1109/TVT.2022.3183596}
}

@ARTICLE{9705497,
  author={Cui, Yaodong and Guo, Aihuang and Song, Chunlin},
  journal={IEEE Wireless Commun. Lett.}, 
  title={TransNet: Full Attention Network for {CSI} Feedback in {FDD} Massive {MIMO} System}, 
  year={2022},
  volume={11},
  number={5},
  pages={903-907},
  doi={10.1109/LWC.2022.3149416}
}

@ARTICLE{9442844,
  author={Zeng, Jun and Sun, Jinlong and Gui, Guan and Adebisi, Bamidele and Ohtsuki, Tomoaki and Gacanin, Haris and Sari, Hikmet},
  journal={IEEE Trans. Cogn. Commun. Netw.},
  title={Downlink {CSI} Feedback Algorithm With Deep Transfer Learning for {FDD} Massive {MIMO} Systems},
  year={2021},
  volume={7},
  number={4},
  pages={1253-1265},
  doi={10.1109/TCCN.2021.3084409}
}

@ARTICLE{10508320,
  author={Liu, Zhenyu and Wang, Li and Xu, Lianming and Ding, Zhi},
  journal={IEEE Trans. Wireless Commun.}, 
  title={Deep Learning for Efficient {CSI} Feedback in Massive {MIMO}: Adapting to New Environments and Small Datasets}, 
  year={2024},
  volume={},
  number={},
  pages={1-1},
  doi={10.1109/TWC.2024.3390583}}

@ARTICLE{10262359,
  author={Li, Xiangyi and Guo, Jiajia and Wen, Chao-Kai and Jin, Shi and Han, Shuangfeng and Wang, Xiaoyun},
  journal={IEEE Trans. Commun.}, 
  title={Multi-Task Learning-Based {CSI} Feedback Design in Multiple Scenarios}, 
  year={2023},
  volume={71},
  number={12},
  pages={7039-7055},
  doi={10.1109/TCOMM.2023.3317924}}

@misc{chen2023csipppnetonesidedoneforalldeep,
      title={{CSI}-{PPPNet}: A One-Sided One-for-All Deep Learning Framework for Massive {MIMO} {CSI} Feedback}, 
      author={Wei Chen and Weixiao Wan and Shiyue Wang and Peng Sun and Geoffrey Ye Li and Bo Ai},
      year={2023},
      eprint={2211.15851},
      archivePrefix={arXiv},
      primaryClass={eess.SP},
      url={https://arxiv.org/abs/2211.15851}, 
}

@inproceedings{rozendaal2021overfitting,
  title={Overfitting for Fun and Profit: Instance-Adaptive Data Compression},
  author={Ties van Rozendaal and Iris {AM} Huijben and Taco Cohen},
  booktitle={Int. Conf. Learn. Represent. (ICLR)},
  year={2021},
  url={https://openreview.net/forum?id=oFp8Mx_V5FL}
}

@article{Yang2021TowardsES,
  title={Towards Empirical Sandwich Bounds on the Rate-Distortion Function},
  author={Yibo Yang and Stephan Mandt},
  journal={ArXiv},
  year={2021},
  volume={abs/2111.12166},
  url={https://api.semanticscholar.org/CorpusID:244527239}
}

@inproceedings{
ballé2018variational,
title={Variational image compression with a scale hyperprior},
author={Johannes Ballé and David Minnen and Saurabh Singh and Sung Jin Hwang and Nick Johnston},
booktitle={Int. Conf. Learn. Represent. (ICLR)},
year={2018},
url={https://openreview.net/forum?id=rkcQFMZRb},
}

@Article{Remcom,

author = {Remcom}, 

title = {{Wireless InSite}}, 

note = {\url{http://www.remcom.com/wireless-insite}.},}

@InProceedings{Alkhateeb2019,
author = {Alkhateeb, A.},
title = {{DeepMIMO}: A Generic Deep Learning Dataset for Millimeter Wave and Massive {MIMO} Applications},
booktitle = {Proc. Inf. Theory Appl. Workshop (ITA)},
year = {2019},
pages = {1-8},
month = {Feb},
address = {San Diego, CA},
}

@book{quinonero2022dataset,
  title={Dataset Shift in Machine Learning},
  author={Qui{\~n}onero-Candela, Joaquin and Sugiyama, Masashi and Schwaighofer, Anton and Lawrence, Neil D},
  year={2022},
  publisher={MIT Press}
}

@inproceedings{ganin2015unsupervised,
  title={Unsupervised Domain Adaptation by Backpropagation},
  author={Ganin, Yaroslav and Lempitsky, Victor},
  booktitle={Int. Conf. Mach. Learn. (ICML)},
  pages={1180--1189},
  year={2015},
  organization={PMLR}
}

@article{zhuang2020comprehensive,
  title={A Comprehensive Survey on Transfer Learning},
  author={Zhuang, Fuzhen and Qi, Zhiyuan and Duan, Keyu and Xi, Dongbo and Zhu, Yongchun and Zhu, Hengshu and Xiong, Hui and He, Qing},
  journal={Proc. IEEE},
  volume={109},
  number={1},
  pages={43--76},
  year={2020},
  publisher={IEEE}
}

@ARTICLE{6758357,
  author={Jaeckel, Stephan and Raschkowski, Leszek and Börner, Kai and Thiele, Lars},
  journal={IEEE Trans. Antennas Propag.},
  title={QuaDRiGa: A 3-D Multi-Cell Channel Model With Time Evolution for Enabling Virtual Field Trials},
  year={2014},
  volume={62},
  number={6},
  pages={3242-3256},
  doi={10.1109/TAP.2014.2310220}
}

@article{DBLP:journals/corr/BengioLC13,
  author = {Yoshua Bengio and Nicholas L{\'{e}}onard and Aaron C. Courville},
  title = {Estimating or Propagating Gradients Through Stochastic Neurons for Conditional Computation},
  journal = {CoRR},
  volume = {abs/1308.3432},
  year = {2013},
  url = {http://arxiv.org/abs/1308.3432},
  eprinttype = {arXiv},
  eprint = {1308.3432},
  note = {Accessed: 2025-01-27}
}

@INPROCEEDINGS{8918798,
  author = {Yang, Qianqian and Mashhadi, Mahdi Boloursaz and Gündüz, Deniz},
  booktitle = {Proc. IEEE 29th Int. Workshop Mach. Learn. Signal Process. (MLSP)},
  title = {Deep Convolutional Compression for Massive {MIMO} {CSI} Feedback},
  year = {2019},
  pages = {1-6},
  doi = {10.1109/MLSP.2019.8918798}
}

@article{DBLP:journals/corr/abs-2102-09270,
  author = {Lucas Theis and Eirikur Agustsson},
  title = {On the Advantages of Stochastic Encoders},
  journal = {CoRR},
  volume = {abs/2102.09270},
  year = {2021},
  url = {https://arxiv.org/abs/2102.09270},
  eprinttype = {arXiv},
  eprint = {2102.09270},
  bibsource = {dblp computer science bibliography, https://dblp.org}
}

@inproceedings{9834372,
  author = {Jankowski, Mikolaj and Gündüz, Deniz and Mikolajczyk, Krystian},
  booktitle = {2022 IEEE Int. Symp. Inf. Theory (ISIT)},
  title = {AirNet: Neural Network Transmission over the Air},
  year = {2022},
  pages = {2451-2456},
  doi = {10.1109/ISIT50566.2022.9834372}
}

@inproceedings{NEURIPS2018_1d94108e,
  author = {Volpi, Riccardo and Namkoong, Hongseok and Sener, Ozan and Duchi, John C and Murino, Vittorio and Savarese, Silvio},
  booktitle = {Adv. Neural Inf. Process. Syst. (NeurIPS)},
  title = {Generalizing to Unseen Domains via Adversarial Data Augmentation},
  editor = {S. Bengio and H. Wallach and H. Larochelle and K. Grauman and N. Cesa-Bianchi and R. Garnett},
  pages = {},
  publisher = {Curran Associates, Inc.},
  url = {https://proceedings.neurips.cc/paper_files/paper/2018/file/1d94108e907bb8311d8802b48fd54b4a-Paper.pdf},
  volume = {31},
  year = {2018}
}

@ARTICLE{10097872,
  author={Zhang, Boyuan and Li, Haozhen and Liang, Xin and Gu, Xinyu and Zhang, Lin},
  journal={IEEE Communications Letters}, 
  title={Model Transmission-Based Online Updating Approach for Massive {MIMO} {CSI} Feedback}, 
  year={2023},
  volume={27},
  number={6},
  pages={1609-1613},
  doi={10.1109/LCOMM.2023.3265680}}

@ARTICLE{10381825,
  author={Zhang, Xudong and Wang, Jintao and Lu, Zhilin and Zhang, Hengyu},
  journal={IEEE Communications Letters}, 
  title={Continuous Online Learning-Based {CSI} Feedback in Massive {MIMO} Systems}, 
  year={2024},
  volume={28},
  number={3},
  pages={557-561},
  doi={10.1109/LCOMM.2024.3350210}}

@ARTICLE{10359472,
  author={Cui, Yiming and Guo, Jiajia and Wen, Chao-Kai and Jin, Shi},
  journal={IEEE Transactions on Wireless Communications}, 
  title={Communication-Efficient Personalized Federated Edge Learning for Massive {MIMO} {CSI} Feedback}, 
  year={2024},
  volume={23},
  number={7},
  pages={7362-7375},
  doi={10.1109/TWC.2023.3339824}}

@ARTICLE{9737435,
  author={Guo, Jiajia and Zuo, Yiping and Wen, Chao-Kai and Jin, Shi},
  journal={IEEE Journal of Selected Topics in Signal Processing}, 
  title={User-Centric Online Gossip Training for Autoencoder-Based {CSI} Feedback}, 
  year={2022},
  volume={16},
  number={3},
  pages={559-572},
  doi={10.1109/JSTSP.2022.3160268}}

@article{Ročková02012018,
author = {Veronika Ročková and Edward I. George and},
title = {The Spike-and-Slab LASSO},
journal = {Journal of the American Statistical Association},
volume = {113},
number = {521},
pages = {431--444},
year = {2018},
publisher = {ASA Website},
doi = {10.1080/01621459.2016.1260469},


URL = { 
    
        https://doi.org/10.1080/01621459.2016.1260469
    
    

},
eprint = { 
    
        https://doi.org/10.1080/01621459.2016.1260469
    
    

}

}

@TECHREPORT{3gpp.38.901,
  organization={3rd Generation Partnership Project ({3GPP})},
  title={{Study on channel model for frequencies from 0.5 to 100 {GHz}}}, 
  institution={3GPP},
  number={TR 38.901},
  version={17.0.0},
  year={2022},
  month={Mar.},
  note={Available: \url{https://www.3gpp.org/ftp/Specs/archive/38_series/38.901/}}
}

@INPROCEEDINGS{11143260,
  author={Sattari, Mehdi and Gündüz, Deniz and Svensson, Tommy},
  booktitle={2025 IEEE 26th International Workshop on Signal Processing and Artificial Intelligence for Wireless Communications (SPAWC)}, 
  title={{Dynamically Fine-Tuned Neural Compressor for {FDD} Massive {MIMO} {CSI} Feedback}}, 
  year={2025},
  volume={},
  number={},
  pages={1-5},
  doi={10.1109/SPAWC66079.2025.11143260}}

@INPROCEEDINGS{10018052,
  author={Jourairi, Oussama and Balcilar, Muhammet and Lambert, Anne and Schnitzler, François},
  booktitle={2022 Picture Coding Symposium (PCS)}, 
  title={{Improving The Reconstruction Quality by Overfitted Decoder Bias in Neural Image Compression}}, 
  year={2022},
  volume={},
  number={},
  pages={61-65},
  doi={10.1109/PCS56426.2022.10018052}}

@INPROCEEDINGS{11106171,
  author={Hübner, Paul and Hu, Andong and Peng, Ivy and Markidis, Stefano},
  booktitle={2025 IEEE International Parallel and Distributed Processing Symposium Workshops (IPDPSW)}, 
  title={{Apple vs. Oranges: Evaluating the Apple Silicon M-Series {SoCs} for {HPC} Performance and Efficiency}}, 
  year={2025},
  volume={},
  number={},
  pages={45-54},
  doi={10.1109/IPDPSW66978.2025.00013}}

@misc{kaplan2020scalinglawsneurallanguage,
      title={Scaling Laws for Neural Language Models}, 
      author={Jared Kaplan and Sam McCandlish and Tom Henighan and Tom B. Brown and Benjamin Chess and Rewon Child and Scott Gray and Alec Radford and Jeffrey Wu and Dario Amodei},
      year={2020},
      eprint={2001.08361},
      archivePrefix={arXiv},
      primaryClass={cs.LG},
      url={https://arxiv.org/abs/2001.08361}, 
}

\end{document}